\documentclass[usenatbib]{mn2e} 
\input psfig.sty 
\usepackage{epsfig}
\usepackage{amsmath,amssymb} \usepackage{psfig}

\newcommand{\kpc} {{\,\rm kpc}} \newcommand{\Gyr} {{\,\rm Gyr}}
\newcommand{\pc} {{\,\rm pc}} 
\newcommand{\kms}{{\,\rm {km\,s^{-1}} }} 

\title[Turbulence in galactic discs] {Large scale galactic turbulence: can self-gravity drive the observed HI velocity dispersions?} \author[Oscar Agertz
     et al.]  {\parbox[t]{\textwidth}{Oscar
   Agertz$^1$\thanks{agertz@physik.unizh.ch}, George Lake$^1$, Romain Teyssier$^{1,2}$, Ben Moore$^1$, Lucio Mayer$^{1,3}$,  \\ Alessandro B. Romeo$^4$}\vspace*{6pt}\\$^1$ Institute for Theoretical Physics, University of Z\"urich, CH-8057 Z\"urich, Switzerland\\$^{2}$ CEA Saclay, DSM/IRFU/SAp, Batiment 709, 91191 Gif-sur-Yvette Cedex, France\\$^3$ Department of Physics, Institute f\"ur Astronomie, ETH Z\"urich, CH-8093 Z\"urich, Switzerland\\$^{4}$ Onsala Space Observatory, Chalmers University of Technology, SE-43992 Onsala, Sweden}
\date{\today}
\begin{document}
\maketitle
\begin{abstract} 
Observations of turbulent velocity dispersions in the HI component of galactic discs show a characteristic floor in galaxies with low star formation rates and within individual galaxies the dispersion profiles decline with radius. We carry out several high resolution adaptive mesh simulations of gaseous discs embedded within dark matter haloes to explore the roles of cooling, star-formation, feedback, shearing motions and baryon fraction in driving turbulent motions. In all simulations the disc slowly cools until gravitational and thermal instabilities give rise to a multiphase medium in which a large population of dense self-gravitating cold clouds are embedded within a warm gaseous phase that forms through shock heating. The diffuse gas is highly turbulent and is an outcome of large scale driving of global non-axisymmetric modes as well as cloud-cloud tidal interactions and merging. At low star-formation rates these processes alone can explain the observed HI velocity dispersion profiles and the characteristic value of $\sim10\kms$ observed within a wide range of disc galaxies. Supernovae feedback creates a significant hot gaseous phase and is an important driver of turbulence in galaxies with a star-formation rate per unit area $\,\gtrsim 10^{-3}\,M_\odot\,{\rm yr}^{-1}\kpc^{-2}$.
\end{abstract}


\section{Introduction}
\label{sect:intro}
\begin{figure}
\psfig{file=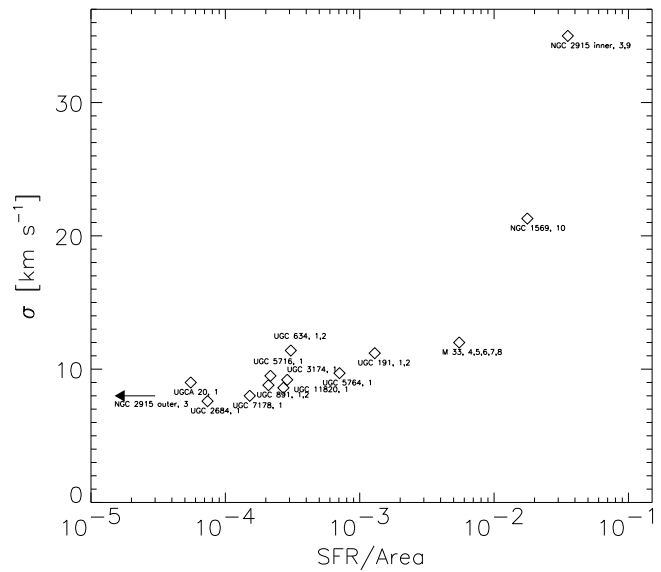,width=255pt}
\caption[]{Characteristic HI gas velocity dispersion of a sample of galaxies as a
function of the derived star formation rate in units $M_\odot\,{\rm
yr}^{-1}\kpc^{-2}$ as plotted by Dib et al. (2006). The figure is
reproduced here by courtesy of Sami Dib, Eric Bell and Andreas Burkert, and by permission of the AAS.}
\label{fig:dib2006}
\end{figure}

The interstellar medium (ISM) is dominated by irregular/turbulent gas motions \citep[e.g.][]{larson81,elmegreen04}. HI emission lines in most spiral galaxies have characteristic velocity dispersions of $\sigma\sim 10 {\rm\, km/s}$ on a scale of a few hundred parsecs, exceeding the values expected from purely thermal effects. The data in Fig.\,\ref{fig:dib2006}, assembled by \cite{dib06}, also shows a transition to much larger values in active/starbursting galaxies. Recent high resolution observations by \cite{petric07} of the nearly face on disc galaxy NGC 1058 \citep[see also][]{dickey90} provides us data on the radial behavior of the vertical velocity dispersion. They find that the dispersion declines with radius from $\sim12-15\,\kms$ in the inner parts to $\sim4-6\,\kms$ in the outer and is uncorrelated with active regions such as star formation sites and spiral arms. This is attributed to small scale ($< 0.7 \kpc$) bulk motions. Petric \& Rupen state that any model attempting to explain turbulence in the ISM \emph{must} also explain the radial decline that also has been detected in previous studies of e.g. NGC 6946 \citep{boulanger92}, NGC 628 \citep{kamphuis93,vanderhulst96}, NGC 2915 \citep{meurer96}.

The main source(s) of energy driving the ISM dynamics is still not clear \citep{burkert06}, even though there are several candidates capable of driving the ISM turbulence \citep{maclow:review04}. A commonly discussed source is of stellar origin i.e. large-scale expanding outflows from high-pressure HII regions \citep{kessel03}, stellar winds or supernovae. Whilst supernovae explosions might dominate the energy input into the ISM \citep[e.g.][]{maclow:review04,dib06}, the mechanism is unable to explain the broad HI lines in galaxies with a low star formation rate (SFR) and in regions of moderate stellar activity as in the outer parts of disc galaxies. Many numerical studies have been carried out to understand the influence of supernovae in galactic discs \citep[e.g.][]{kim01,deavillez04,deavillez05,slyz05,maclow05,joungmaclow06}. \cite{dib06} reproduced the starbursting transition seen in Fig.\,\ref{fig:dib2006} but was unable to produce velocity dispersions larger than $\sim 3\kms$ for low values of SFR/Area. This strongly suggests that something else is contributing to the energy budget. In addition, large scale holes, usually attributed to correlated supernovae explosions \citep[e.g.][]{puche92}, are in some cases surprisingly \emph{uncorrelated} to stellar activity \citep{rhode99}.

Another source of turbulence is galactic rotation. This is a huge reservoir of energy \citep{fleck81} and any mechanism able to generate random motions from ordered circular motion could sustain turbulence for many orbital times. Numerical work of \cite{wada02} and \cite{wada07} has shown that realistic global models of galactic discs form a very complicated turbulent velocity field associated with a multiphase ISM. The only active source for this is shear coupled to gravitational and thermal instability. Local isothermal simulations of the ISM done by \cite{kimostriker07} (also previous work e.g. \cite{kimostriker01} and \cite{kimostriker03}) support this notion. They demonstrated that gas in a marginally stable galactic discs obtains, under certain conditions, velocity dispersions as large as the sound speed (here $c_s=7\kms$) due the swing-amplifier \citep{goldreichlyndenbell65b,juliantoomre66,toomre81,fuchs01}. The swing-amplifier is when a leading wave is amplified into a trailing wave. The underlying mechanism is shear and self-gravity.

\cite{fukunagatosa89} showed that rotational energy randomizes the motions of the cold cloud component of a galactic disc via gravitational scattering from their random epicyclic motions. This was was later quantified by \cite{gammie91} who showed that the cloud velocity dispersion could reach $\sim 5-6\kms$ in this way, in agreement with observations \citep{starkbrand89}. We will discuss this mechanism and its impact on the ISM in more detail in Sect.\ref{sect:results}. 

The Magneto-Rotational-Instability (MRI) \citep{balbushawley91,sellwoodbalbus99} coupled with galactic shear
is also a possible driver of turbulence. \cite{piontekostriker04} and \cite{piontekostriker05} obtained reasonable values of $\sim 8 \kms$ under favorable conditions. This mechanism becomes significant at low densities and might be important in the more diffuse outer part of galaxies.

In this paper we carry out high-resolution 3-dimensional Adaptive Mesh Refinement (AMR) simulations to form a realistic multiphase ISM in which we can disentangle the contributing effects of self-gravity and supernovae driven turbulence. The simulations incorporate realistic prescriptions for cooling, star formation and supernovae feedback. Similar numerical simulations have been carried out before \citep[e.g.][]{gerritsenicke97,wada02,bottema03,taskerbryan06,wada07} but without addressing directly the issues discussed in this paper. 

The paper is organized as follows. In Sect.\,\ref{sect:nummod} we describe the numerical method used for this work and the setup of the galactic discs. In Sect.\,\ref{sect:results} we present the results from the numerical simulations, where the results treating the turbulent ISM are given in Sect.\ref{sect:turb}. Sect.\,\ref{sect:conclusions} summarizes and discusses our conclusions.

\section{Numerical modelling}
\label{sect:nummod}
\subsection{The code and subgrid modelling}
\label{sect:code}
We use the adaptive mesh refinement (AMR)
hydrodynamics code RAMSES \citep{teyssier02}. The code uses a second order
Godunov scheme to solve the Euler equations. The equation of state of the gas is that of a perfect mono-atomic gas with an adiabatic index $\gamma=5/3$. Self-gravity of the gas is calculated by solving the Poisson equation using the multigrid method \citep{brandt77} on the 
coarse grid and by the conjugate gradient method on finer ones. The collisionless star particles are evolved using the particle-mesh technique. The dark matter 
is treated as a smooth background density field that is added as a static source term in the Poisson solver. The code adopts the cooling function
of \cite{sutherlanddopita93}  for cooling at temperatures $10^4-10^{8.5}\,$K. We extend cooling down to 300 K using the parametrization of \cite{rosenbregman95}. The effect of 
metallicity is approximated by using a linear scaling of the functions. 

The star formation recipe is described in \cite{dubois08} but we summarize the main points here for completeness. In a cell, gas is converted to a star particle using a Schmidt law
\begin{equation}
\dot{\rho}_*=-\frac{\rho}{t_*}\, \rm{if} \,\rho>\rho_0 \qquad \dot{\rho}_*=0\,\,\rm{otherwise},
\end{equation}
where $t_*$ is the star formation time scale and  $\rho_0$ is an arbitrary threshold that should be chosen to carefully make physical sense when related to the resolution and cooling floor. The star formation timescale is related to the local free-fall time,
\begin{equation}
t_*=t_0\left(\frac{\rho}{\rho_0}\right)^{-1/2}.
\end{equation}
The parameters $\rho_0$ and $t_0$ are in reality scale dependent and not very well understood theoretically. A common way to get around this is to calibrate them to star formation rates in local galaxies i.e. to the \cite{kennicutt98} law and make sure that the values are compatible with modern estimates of star formation efficiencies \citep{krumholztan05} of $\sim 1-2\,\%$ per free-fall time in giant molecular clouds (GMCs). For example, if the star formation threshold $\rho_0=100\,\rm{cm}^{-3}$, the free fall time is $5\,$Myr meaning we can use $t_0=250\,$Myr to get $2\,\%$ efficiency per free fall time. As soon as a cell is eligible for star formation, particles are spawned using a Poisson process where the stellar mass is connected to the chosen threshold and code resolution \citep[see][]{dubois08}. 

The implementation of supernovae feedback is also described in the above reference (see their Appendix A). In the simulations that include feedback we assume that $50\,\%$ of the total supernovae energy, $E_{SN}=10^{51}$ ergs, goes into thermal energy where $\eta_{\rm SN}=10\,\%$ of each solar mass of stars that is formed is recycled as supernovae ejecta. The energy and gas release is also delayed by $10\,$Myr from the time of explosion by creating debris particles on at the time of explosion. By delaying the energy and mass release we allow for it to take place outside of dense environments, hence preventing it from radiating away too quickly. We follow the prescription of \cite{dubois08} and apply a large mass loading factor for the debris particles, i.e. $\eta_W=1.0$. We use thermal rather than kinetic energy releases since we resolve the clumpy ISM and follow shocks self-consistently. In addition, a model that allows for debris particles to transfer kinetic energy is no longer valid as a Sedov explosion assumes a homogeneous medium to propagate into. While any treatment is inherently sub-grid, the supernovae impact should converge with enough resolution \citep{ceverino08}.

In order to model subgrid gaseous equation of states and to avoid artificial gas fragmentation, the gas is given a polytropic equation of state as it crosses $\rho_0$. The temperature is set to
\begin{equation} 
T=T_0\left(\frac{\rho}{\rho_0}\right)^{\gamma_0-1}.
\end{equation}
$T_0$ is set to be the cooling floor of our simulations for consistency and $\gamma_0=2.0$. 

\subsection{Initial conditions}
\begin{figure*}
\begin{tabular}{cccc}
\psfig{file=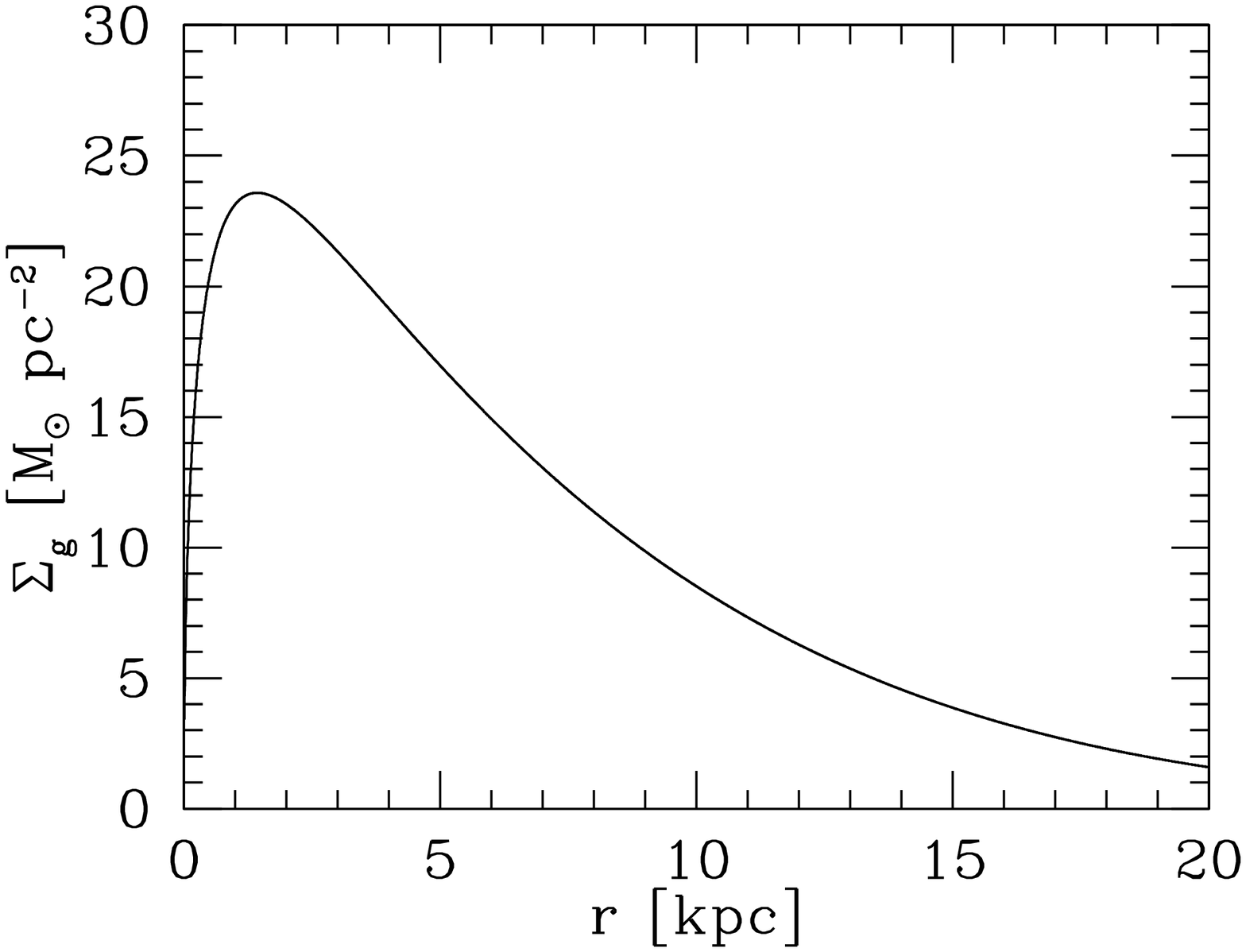,width=116pt} &
\psfig{file=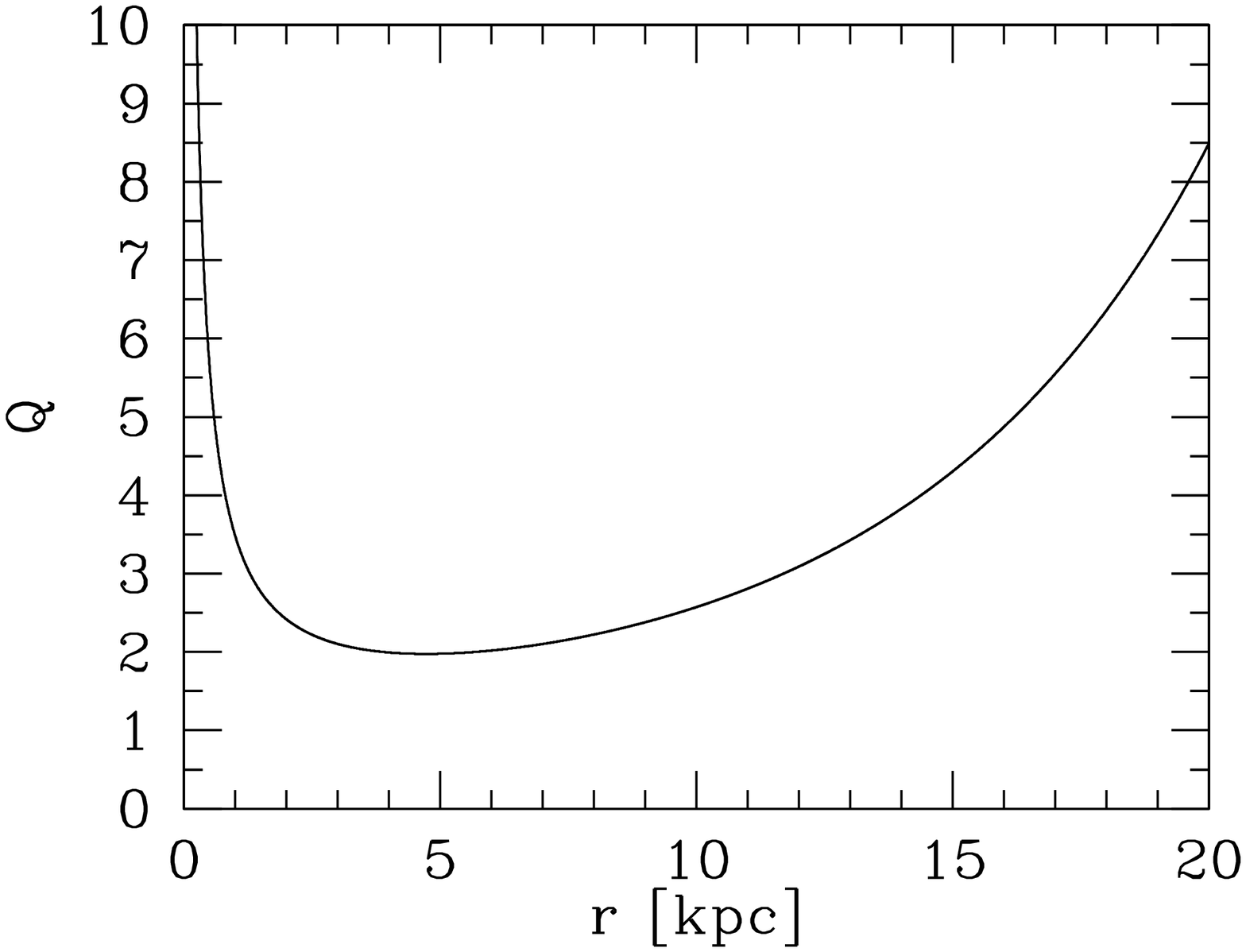,width=116pt} &
\psfig{file=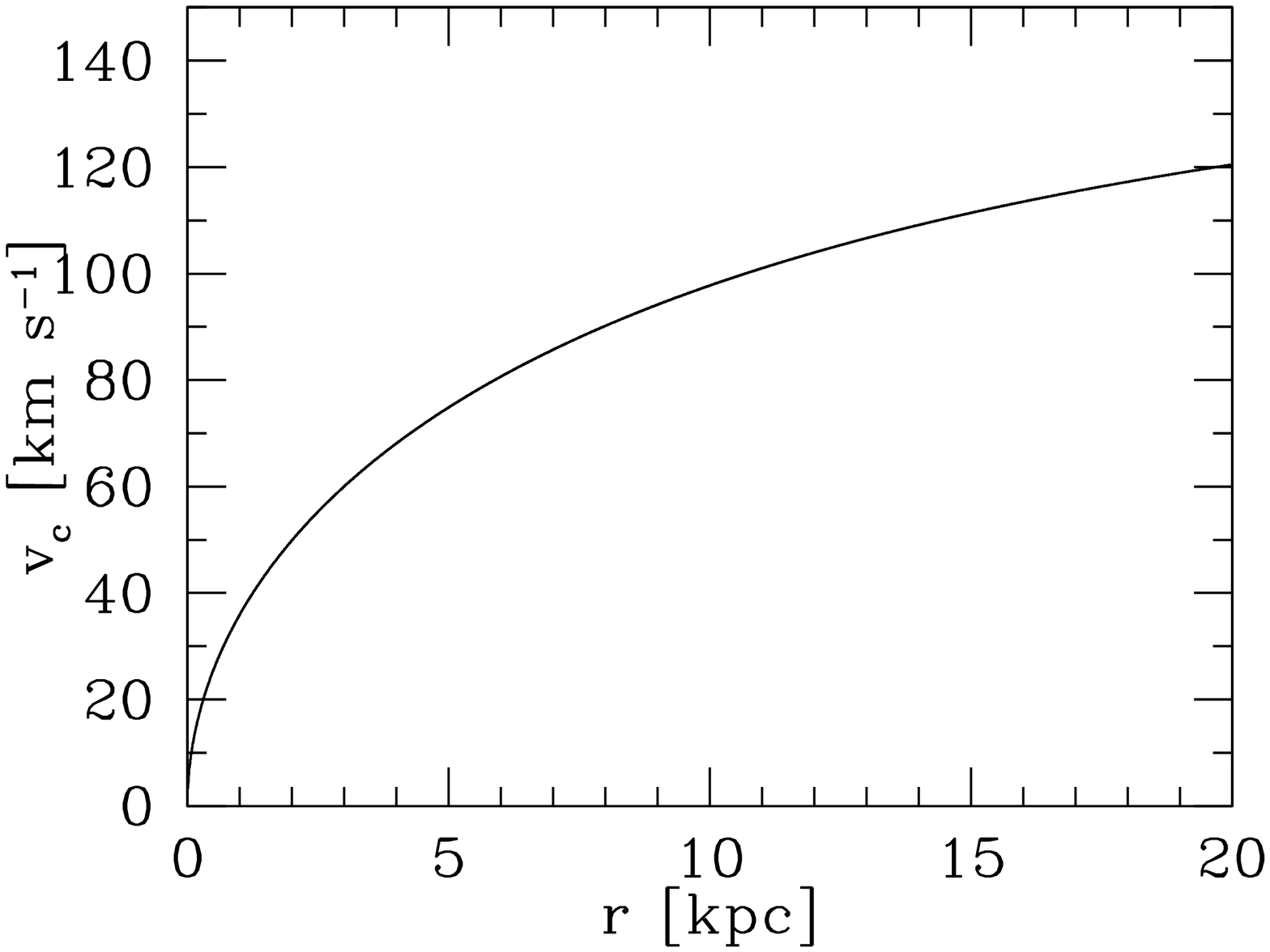,width=116pt} &
\psfig{file=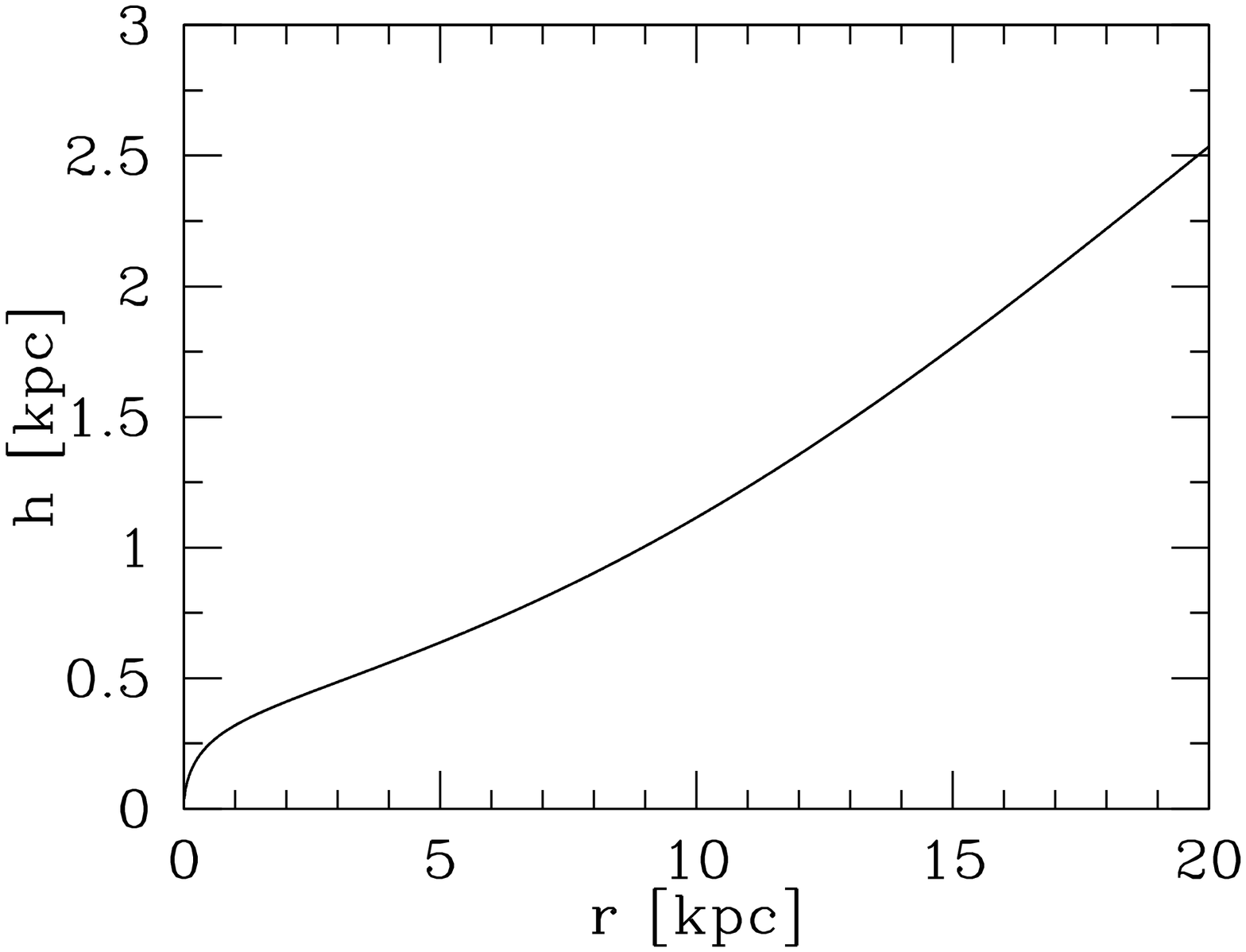,width=116pt} \\
\end{tabular}
\caption[]{Characteristics of the initial gas distribution in all simulations, apart from RUN6 and RUN8. The panels show, from left to right, the radial dependence of the surface density, the gaseous $Q$-value, the circular velocity and the scale height.}
\label{fig:IC}
\end{figure*}
Our initial condition (IC) is an axisymmetric galactic gas disc in equilibrium with an NFW \citep{nfw97} dark matter halo. All relevant IC characteristics are presented in Fig. (\ref{fig:IC}). The disc is initially isothermal at $T=10^4$ K having an exponential density profile, in cylindrical coordinates $r$ and $z$,
\begin{equation}
\rho(r,z)={\rm sech}^2(z/h(r))\rho_0e^{-r/r_{\rm 0}},
\end{equation}
where $r_{\rm 0}$ is the scale radius and $h(r)$ the scale height. The $\rm{ sech}^2$-term owes to the isothermality of the gas such that 
\begin{equation}
h(r)=\frac{c_s^2}{\pi G\Sigma(r)}
\end{equation}
where $c_s$ is the local sound speed and $\Sigma (r)$ the local total
surface density ($\Sigma=\Sigma_{\rm gas}+\Sigma_{\rm DM}$), naturally leading to a flaring disc.
Experiments using a radial $1/r$ gas distribution were also performed
(not discussed further) without any significant differences. 

We choose to model an M33 type galactic disc as it is a nearby well observed gas rich system. All global characteristics of the initial disc are in agreement with the observations presented by \cite{corbelli03}. M33 has a total gas mass (HI + HII + He)
of $\sim 3.2\times 10^9 M_\odot$ and an estimated stellar mass of $3-6\times 10^9 M_\odot$. We choose the initial gas mass to
be in the high end of the total baryonic mass, i.e. $\approx 9.2\times10^9 M_\odot$ as a lot of
the outer material will not be a part of what we would associate with
a galaxy and is only an artifact of the way we model isolated
discs. The gas is assumed to have a mean metallicity of $0.3\,Z_{\odot}$.

We initialize the disc in a stable configuration where most of the disc has a Toomre parameter $Q\sim 2-3$. These large values of $Q$
are desirable as we want the cooling to initiate instabilities and not
our choice of initial conditions. A fairly large scale radius of
$r=4.0\kpc$ is used. As this only reflects the
very early setup of a forming disc galaxy this will not be an issue in
our modelling. The dark matter halo has a concentration of $c=8.0$,
scale radius $R_s=35.0\kpc$ and a total mass of $10^{12} M_\odot$. These model parameters can be perceived as odd but is necessary for a best fit NFW-halo which, in accordance with observations \citep{corbelli03}, reproduce a dark halo mass that within $17\,\kpc$ is $\sim 5\times10^{10}\,M_\odot$. This mass sets a lower limit on the actual dark matter halo mass. The HI velocity profile is still (slowly) rising at this radius. Scenario with different mass profiles, gas masses, shear and cooling floors are also explored. 

Numerically this setup is initialized at a resolution of $100\pc$
using a nested hierarchy of grids situated in a simulations cube of
size $L_{\rm box} = 200\kpc$. We achieve higher resolution by refining
cells both on based on a density and Jeans mass criterion (see
Sect.\,\ref{sect:numconc} for details). The maximum allowed resolution is
indicated in Table 1.  

\subsection{Simulation suite}
\begin{table*}
\caption{Performed simulations.}
\label{table:simsummary}
\begin{minipage}{140mm}
\begin{tabular}{lcccc}\\
\hline \hline Simulation & Min. $\Delta x$ & $\rho_0$ [cm$^{-3}$] & Cooling floor, $T_{\rm 0}$ & Modelling comments \\ \hline \hline
RUN0 & 24 pc & -- & 300 K & Hydrodynamics with self-gravity\\ 
\hline
RUN1 & 24 pc & 10 & 300 K & Like RUN0 + star formation\\ 
RUN2 & 6 pc & 100 & 300 K & Like RUN1\\
\hline  
RUN3 & 24 pc & 10 & 300 K & Like RUN1 + supernova feedback\\
\hline 
RUN4 & 24 pc & -- & $1\,000$ K & Like RUN0\\ 
RUN5 & 24 pc & -- & $10\,000$ K & Like RUN0\\
RUN6 & 24 pc & 10 & 300 K & Like RUN1 but with 1/3 of
gas mass\\ 
RUN7 & 24 pc & -- & $5\,000$ K & Like RUN0 \\ 
RUN8 & 24 pc & 10 & 300 K & Like RUN1 but higher halo concentration\\ 
\hline\hline
\end{tabular}
\end{minipage}
\end{table*}

The performed runs are listed in Table 1. RUN0 serves as our base run
where we only consider the self-gravitating cooling gas and dark matter. RUN1
introduces star formation as does RUN2 but at a higher
resolution. RUN3 is identical to RUN1 but implements the feedback
prescription described in Sect.\,\ref{sect:code}. These are our four
fiducial simulations to understand the importance of these physical mechanisms. To explore how choices of the gas cooling changes the outcome, RUN4, 5 and 7 are identical to RUN0 except for a truncation in the cooling function at the indicated thresholds. This will determine the ability of a disc to develop a gravitoturbulent state \cite[see e.g.][]{gammie01}. RUN6 adopts 1/3 of the gas mass, making it a much more stable
system. Finally, RUN8 adopts a very concentrated halo ($c=40$,
$R_s=7\kpc$, $M=3\times10^{11} M_\odot$) peaking at $120\kms$ to assess the
influence of a different shear, $d\Omega/dR$. 

\subsection{Numerical considerations}
\label{sect:numconc}
In these types of experiments it is important to consistently
resolve the Jeans scale associated with the chosen cooling floor. \cite{truelove97} demonstrated, using isothermal simulations, that at least 4 resolution elements are
necessary to avoid artificial gas fragmentation. Our strategy is to choose realistic star formation density thresholds that together with the cooling temperature floor  gives us a Jeans scale that can be resolved according to the Truelove criteria. The Jeans length given by
\begin{equation}
\lambda_{\rm J}=\sqrt{\frac{\pi c_{\rm s}^2}{G\rho}}.
\end{equation}
In our simulations, where the temperature floor is set at $T=300\,$K, this can be rewritten as 
\begin{equation}
\label{eq:jeans}
\lambda_{\rm J}\approx 312\sqrt{1/n} \pc
\end{equation}
where $n$ is expressed in ${\rm cm}^{-3}$. In order to satisfy the Truelove criterion we adaptively refine on a Jeans mass down to a resolution of $\Delta x=24 (6)\pc$ in RUN1 (RUN2) where we have set the star formation density threshold to $\rho_0=10 (100)\,{\rm cm}^{-3}$. In addition to this precaution, the background ISM polytropic EOS (see Sect.\,\ref{sect:code}) is activated at the same threshold, ensuring us that the Jeans scale never falls below the minimum value of $\sim 100(25)\pc$ set by Eq.\,\ref{eq:jeans} (at $n=10(100)\,{\rm cm}^{-3}$). Additional simulations have been conducted adopting 16 cells per Jeans length to asses the fidelity of the gravitational fragmentation without any significant difference in outcome. No physical perturbations for gravitational instability are seeded in the initially smooth disc meaning the actual perturbations existing arise from the AMR grid. As we are only interested in the long-time ($t>1\Gyr$) dynamical evolution of the system, the actual morphology of the early unstable disc is of little importance. 
\section{Results}
\label{sect:results}
As we will describe in Sect.\,\ref{sect:turb}, the response to gravitational instabilities, both in the form of bound structures and local non-axissymmetric instabilities, is an important source of turbulence. Therefore, before addressing the issue of turbulence we characterize the global gas evolution, phase-structure (density and temperature) and stability of the simulated galactic discs in Sect.\,\ref{sect:general} and Sect.\,\ref{sect:composition}. 
\begin{figure*}
\begin{tabular}{cccc}
\psfig{file=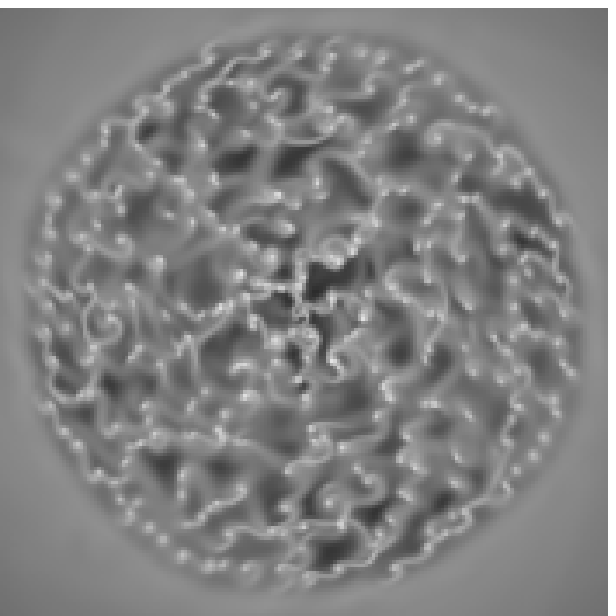,width=115pt} &
\psfig{file=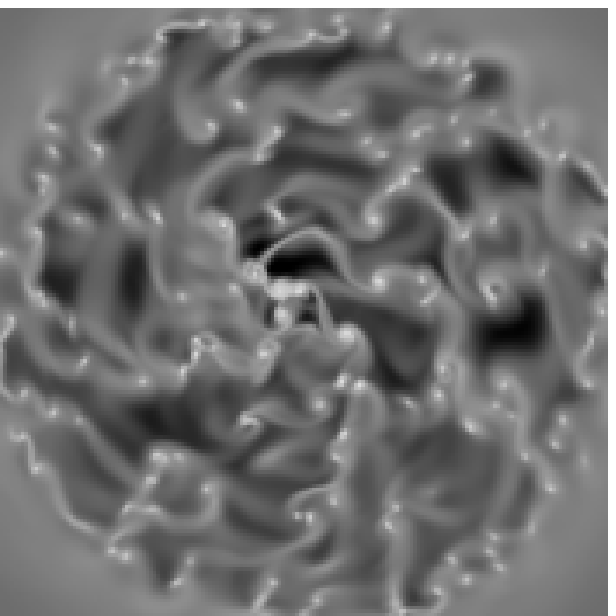,width=115pt} &
\psfig{file=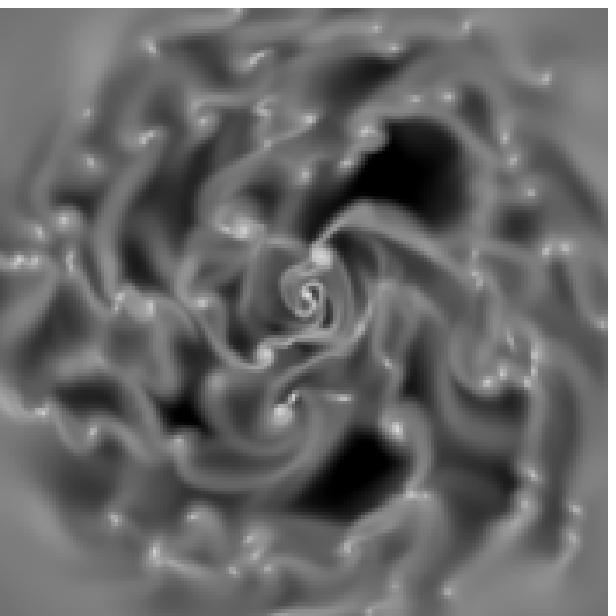,width=115pt} &
\psfig{file=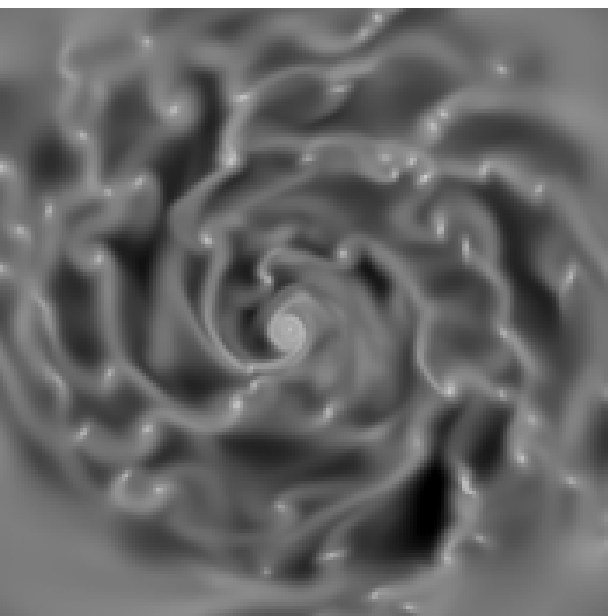,width=115pt} \\
\psfig{file=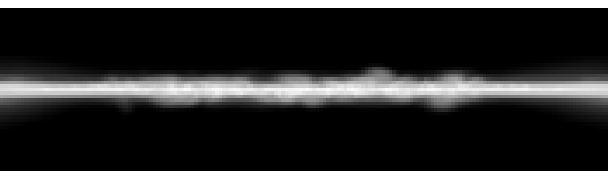,width=115pt} &
\psfig{file=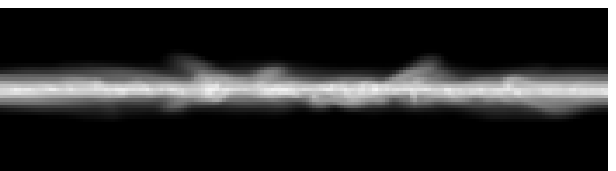,width=115pt} &
\psfig{file=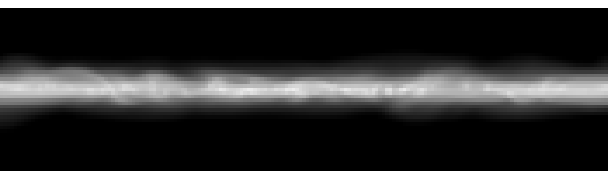,width=115pt} &
\psfig{file=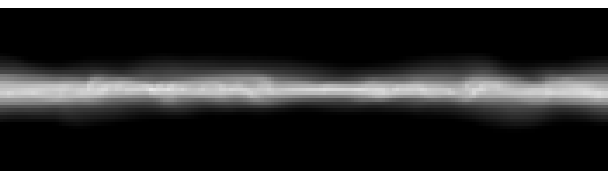,width=115pt} \\
\psfig{file=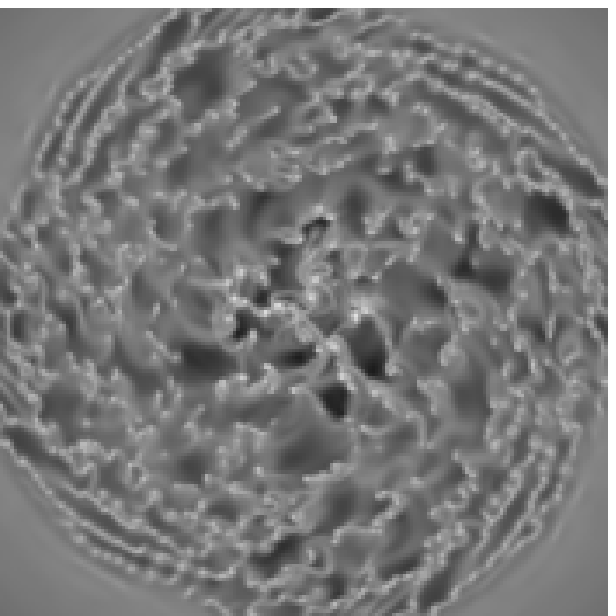,width=115pt} &
\psfig{file=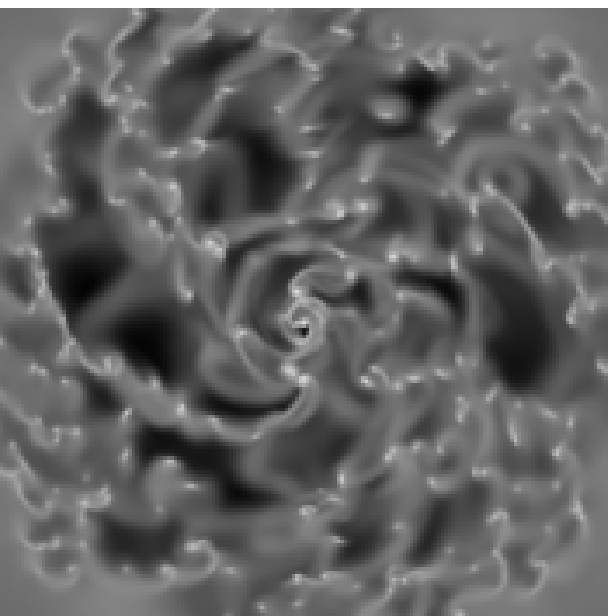,width=115pt} &
\psfig{file=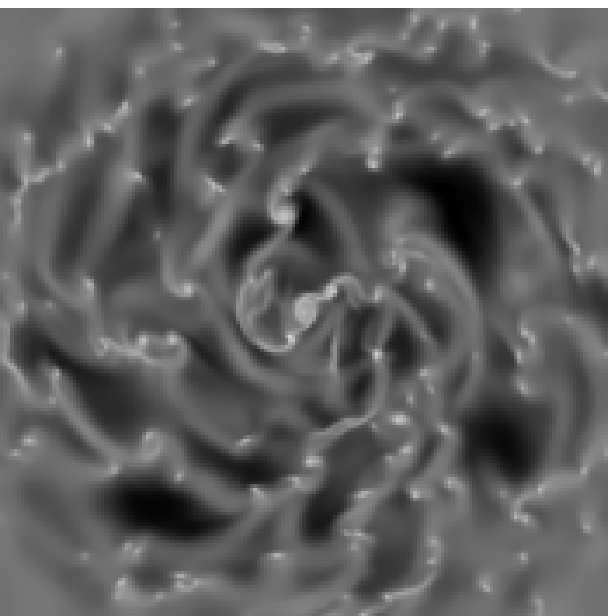,width=115pt} &
\psfig{file=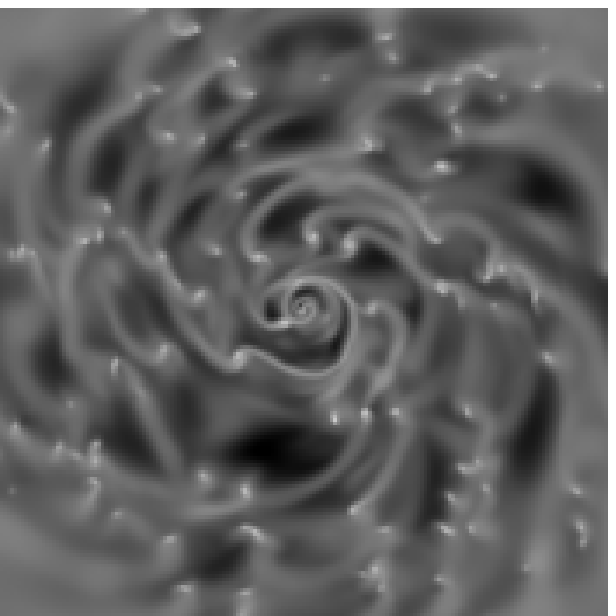,width=115pt} \\
\psfig{file=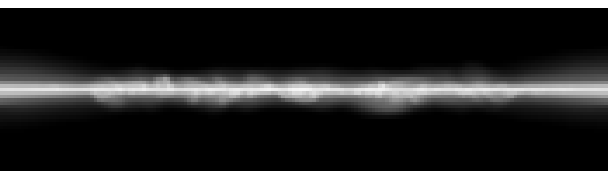,width=115pt} &
\psfig{file=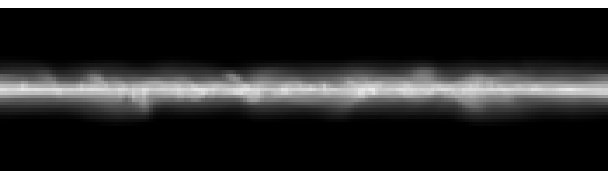,width=115pt} &
\psfig{file=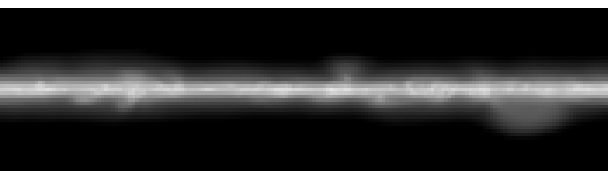,width=115pt} &
\psfig{file=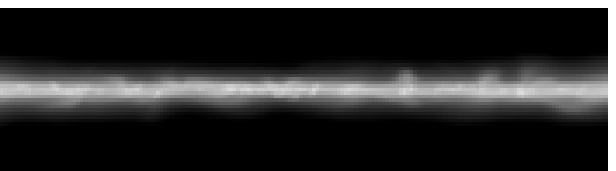,width=115pt} \\
\psfig{file=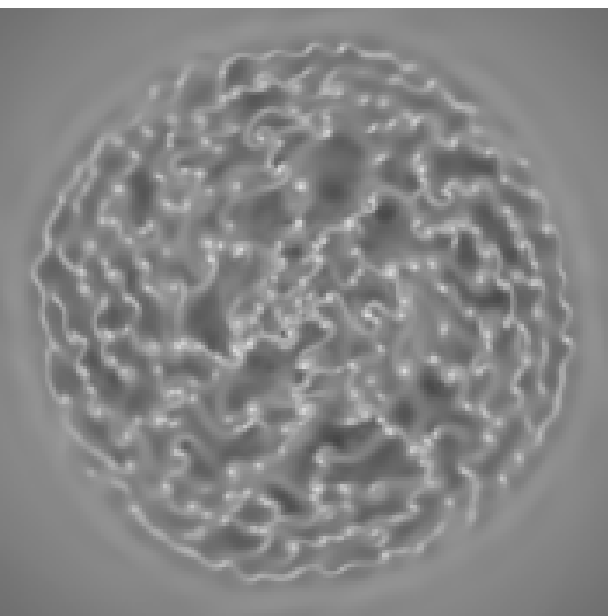,width=115pt} &
\psfig{file=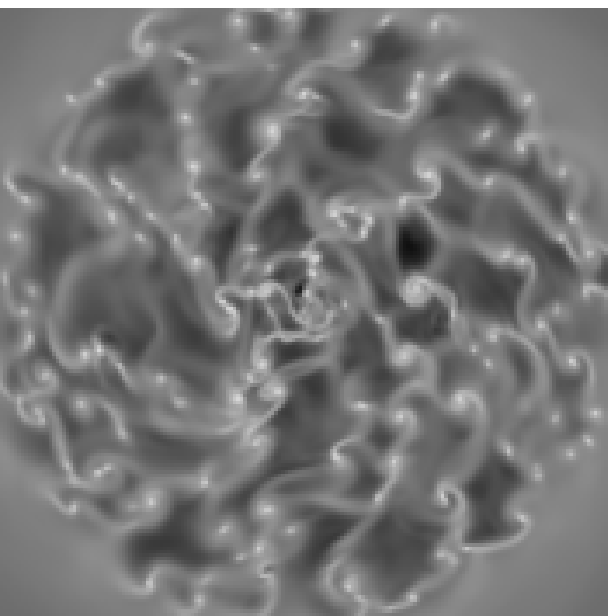,width=115pt} &
\psfig{file=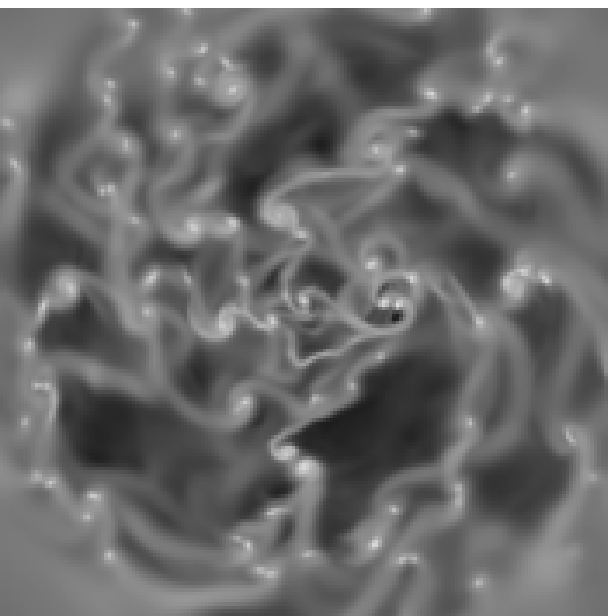,width=115pt} &
\psfig{file=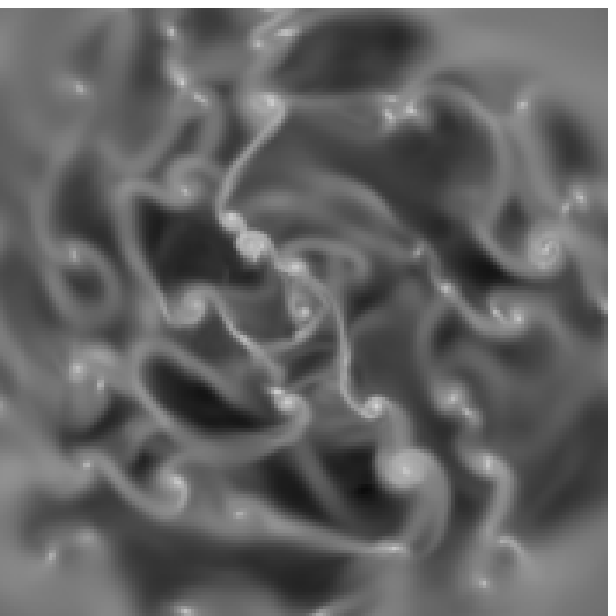,width=115pt} \\
\psfig{file=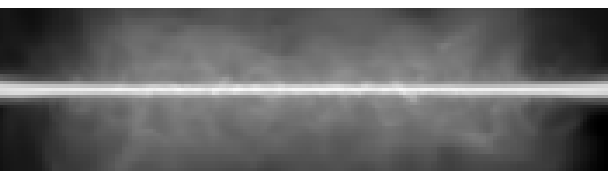,width=115pt} &
\psfig{file=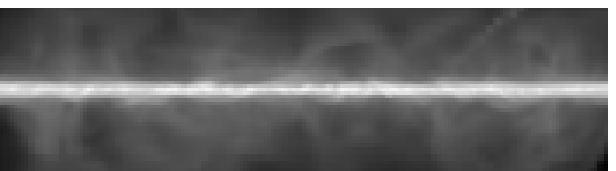,width=115pt} &
\psfig{file=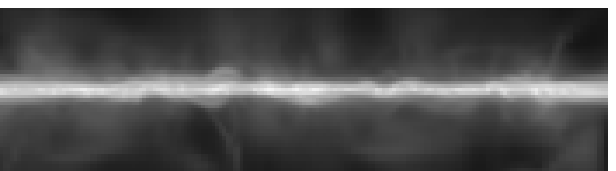,width=115pt} &
\psfig{file=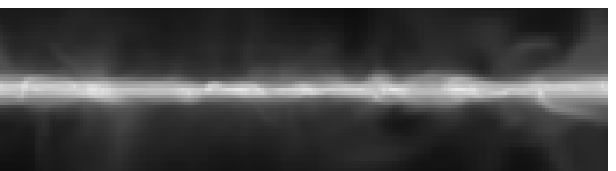,width=115pt} \\
\end{tabular}
\caption[]{Logarithmic column density plots of the gas in the range $\Sigma_{\rm g}=10^{18}-10^{23}\,\rm{cm}^{-3}$. Each panel shows a face-on $30\times30\,\kpc^2$ map centered on
 the disc. The associated edge-on map is 8 kpc in
 height. From top to bottom we see RUN1, RUN2 and
 RUN3 at times, from left to right, $t=0.5, 1.0, 1.5$ and $2.0\Gyr$.}
\label{fig:maps}
\end{figure*}
\subsection{General evolution and morphology}
\label{sect:general}
\subsubsection{Gas evolution}
\label{sect:evol}
\begin{figure*}
\begin{tabular}{ccc}
\psfig{file=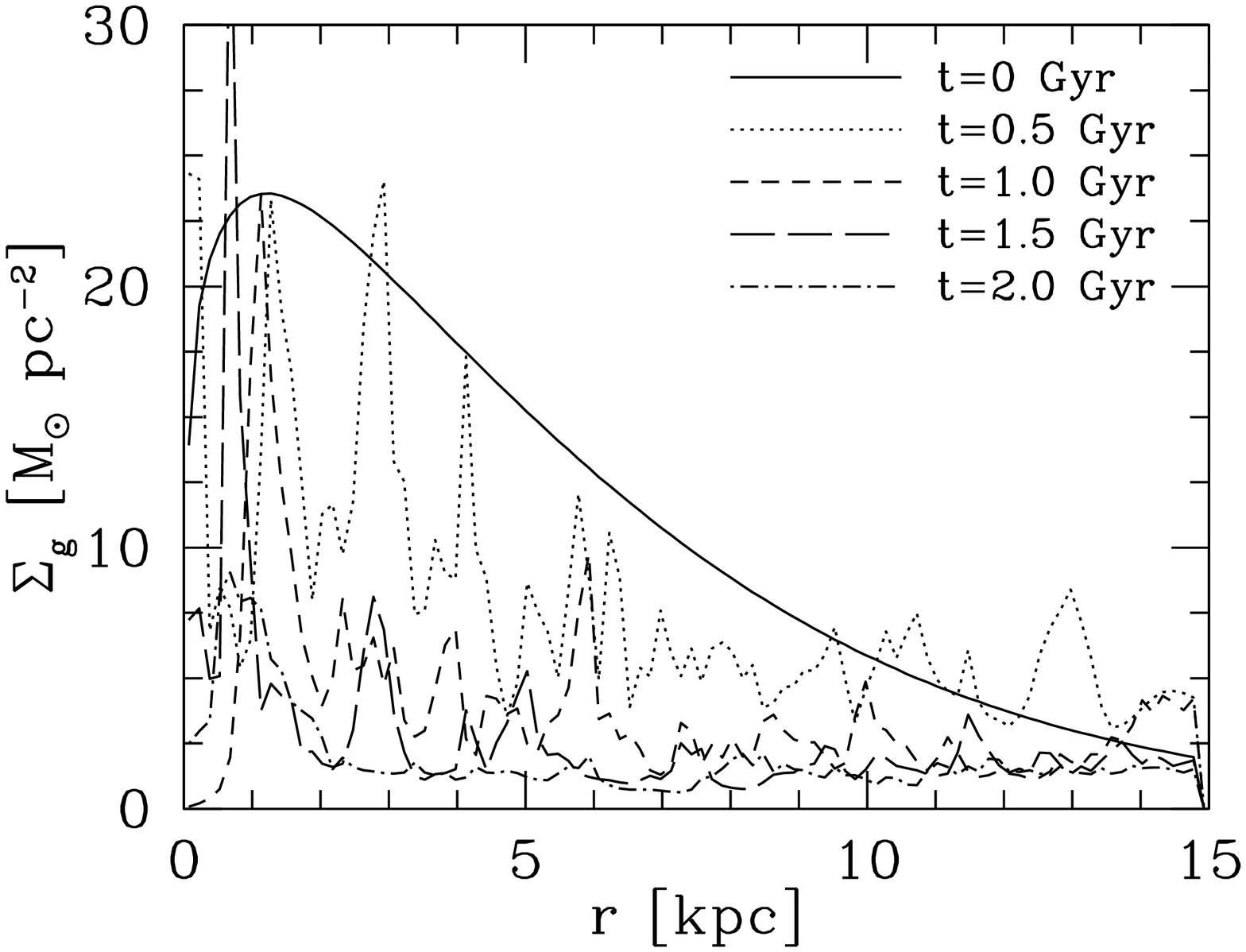,width=161pt} &
\psfig{file=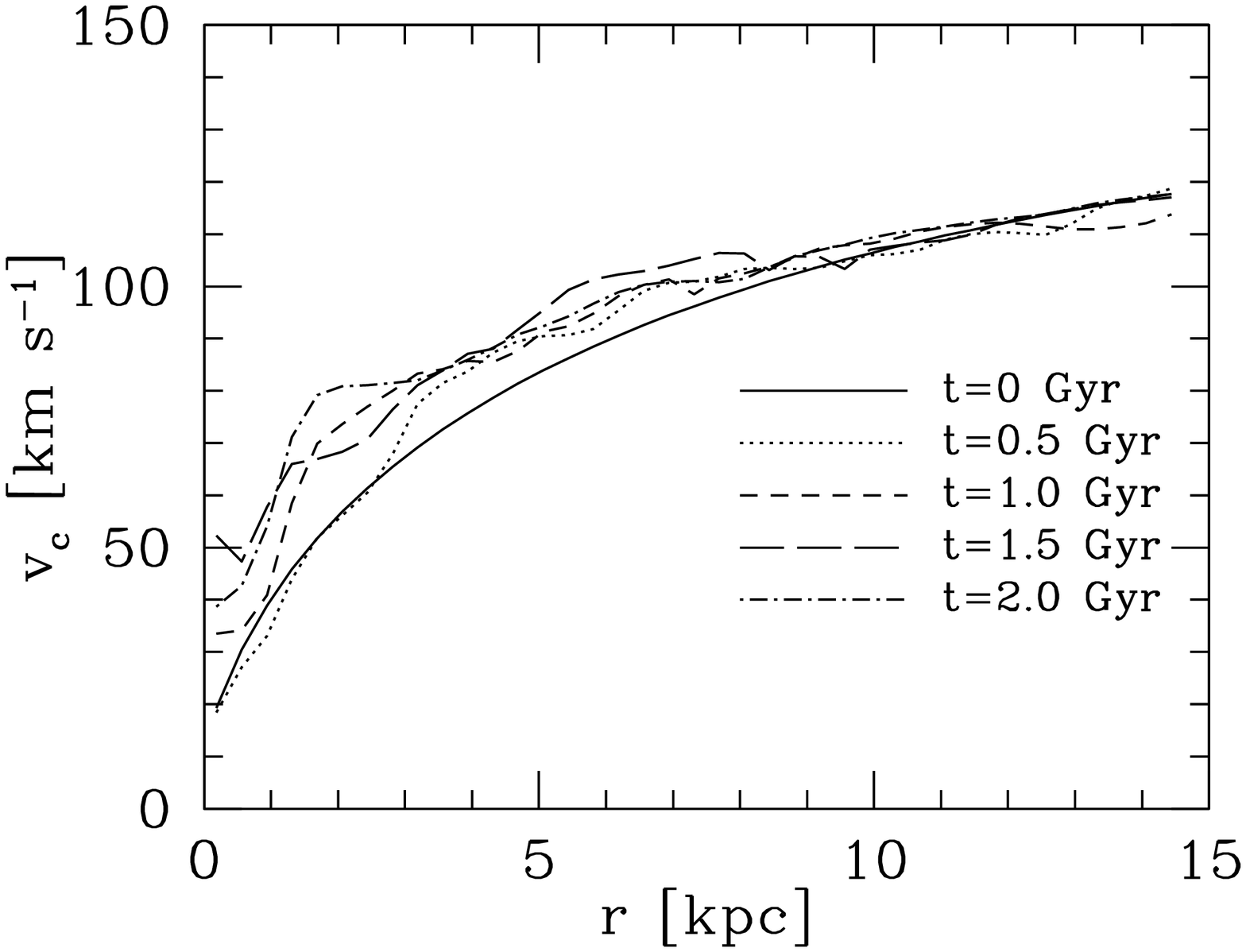,width=161pt} &
\psfig{file=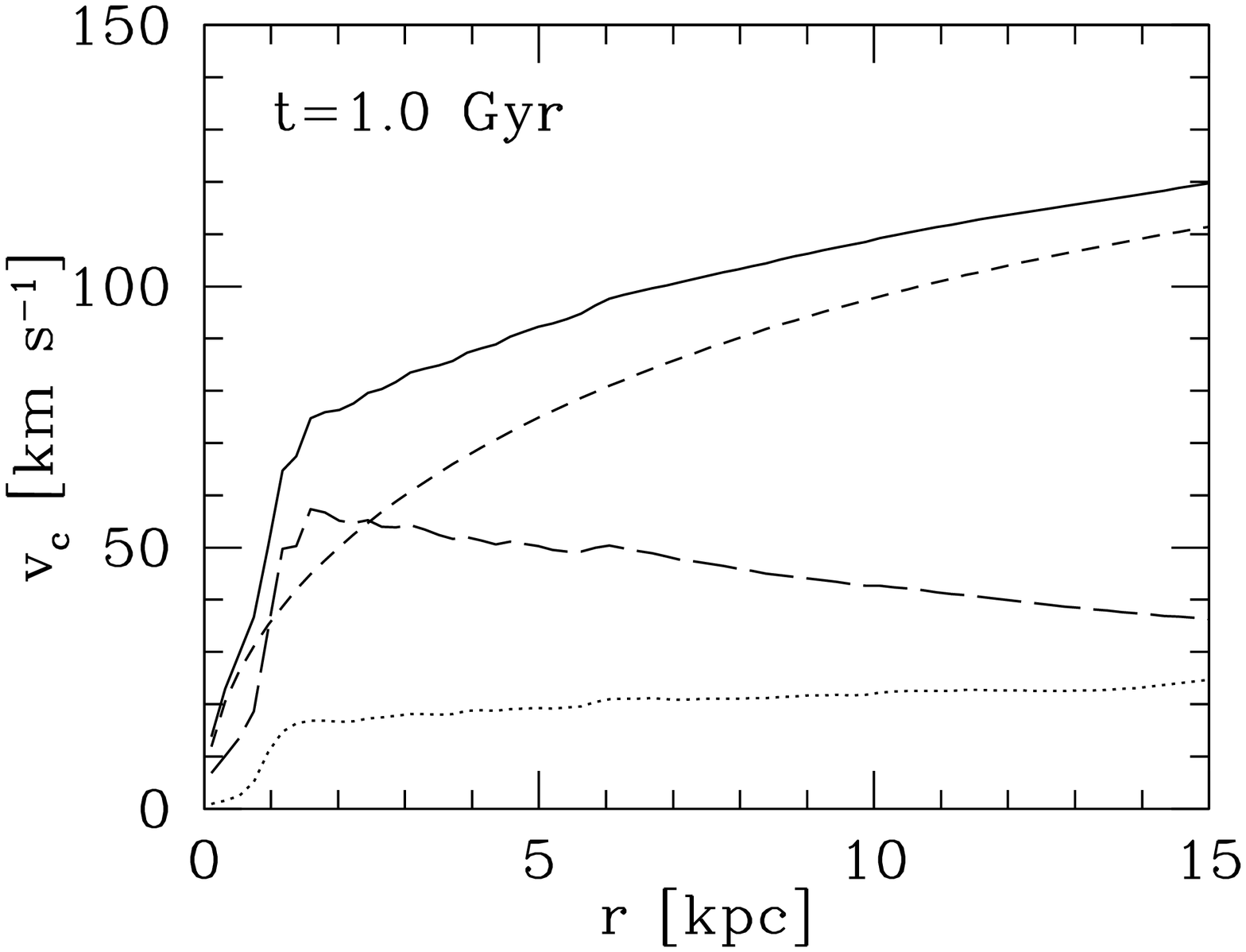,width=161pt} \\
\end{tabular}
\caption[]{Time evolution of the surface density (\emph{left}) and rotational velocity (\emph{middle}) for the gas component in RUN1. The contributions to the rotational velocity (solid line) at $t=1.0\Gyr$ (\emph{right}) from the gaseous (dotted line), stellar (long-dashed line) and dark matter (dash line) components are in good agreement with observations of M33.}
\label{fig:statevol}
\end{figure*}
\begin{figure}
\center
\psfig{file=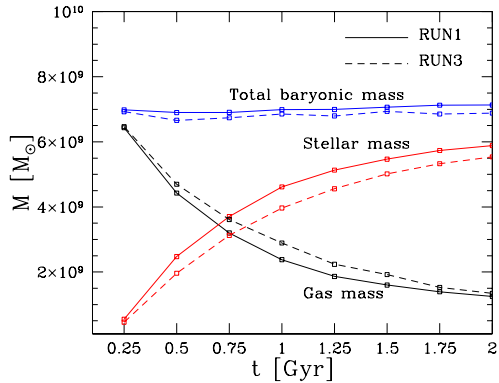,width=200pt}
\caption[]{Time evolution of the mass of the gas and stellar component in a $30\kpc$ cube centered on the disc, as seen in Fig\,\ref{fig:maps}.}
\label{fig:totmass}
\end{figure}

Fig. \ref{fig:maps} shows the total gas surface density maps of RUN1, RUN2 and RUN3 at $t=0.5,1.0,1.5$ and $2.0$ Gyr. All simulations evolve in a similar fashion: the initial gas distribution cools down slowly, loses pressure support and contracts in the vertical direction. After a few 100 Myr the central part of the disc is cold enough to become gravitational and thermally unstable and fragments into bound clouds. This process quickly proceeds to larger radii. Non-axissymmetric instabilities such as swing amplification aids the process everywhere, especially in the outer parts where the gas is only mildly unstable. The formation of bound clouds and elongated structures such as shearing filaments is very similar to that found by e.g. \cite{kimostriker03} and \cite{kimostriker07} for an unstable or marginally stable ISM. This clumpy structure is also visible in the evolution of the total gas surface density and rotational velocity in Fig.\,\ref{fig:statevol}. The decrease of mean surface density is due to star formation. Fig.\,\ref{fig:statevol} also shows the contribution to the rotational velocity at $t=1.0\Gyr$ from gas, stars and dark matter. We can clearly see that the initially gas and dark matter only system has evolved to a state in which the relative contributions and their magnitudes agree well with M33 observations \citep[e.g][]{corbelli03}. The evolution of the total gas and stellar mass in RUN1 and RUN3 are shown in Fig.\,\ref{fig:totmass}. We note that a long-time evolution ($t>1.5\Gyr$) of the galactic discs will force them to move away from a gas rich system such as M33, having $\sim 30\,\%-50\,\%$ of its baryonic mass in gas, approaching $\sim 20\,\%-22\,\%$ at $t= 2\Gyr$. Taking gas infall into account and using a more realistic star formation prescription could remedy this.

We observe significant cloud-cloud and cloud-ISM interactions as the disc evolves. The clouds undergo both collisions leading to coalescence as well as tidal and long range interactions inducing torques into the gas. Shearing wavelets form out of the disc in between the cold clouds. These structures interact with each other as well as the clouds for the entire simulation period. The clumpy ISM acts as an effective viscosity \citep{linpringle87} forcing material to sink to the center whilst smaller clumps stay on more regular orbits at larger radii. This effect is stronger at early times when the typical cloud collision times-scale is short. In between dense clouds and stretched filaments, the ISM also develops under-dense regions ($\Sigma \lesssim10^{18}$cm$^{-2}$) on scales of $500\pc$ to several kpc. At later times, signatures of large scale spiral structure appear in the gaseous disc in which the cold clouds align. 

In what way do the simulations differ? RUN1 and RUN2 evolve in a very similar fashion. However, the higher resolution in RUN2 means that further swing amplified instabilities can occur in the outer parts (see $t=0.5$ Gyr in Fig. \ref{fig:maps}). Apart from this, the overall morphology and statistics are in good agreement throughout the whole simulation time, indicating convergence. The feedback in RUN3 successfully ejects hot low density gas into the ISM as well as out of the disc plane. This process is very efficient at early times when the SFR/Area is high but calms down as the SFR self-regulates, see Sect.\,\ref{sect:feedback} for discussion. The feedback also alters the structure of the disc. As seen in Fig.\,\ref{fig:maps}, the late time spiral patterns are not as pronounced as in RUN1 and RUN2 and fewer low-mass clouds have survived.

\subsubsection{Star formation}
\begin{figure*}
\begin{tabular}{cc}
\psfig{file=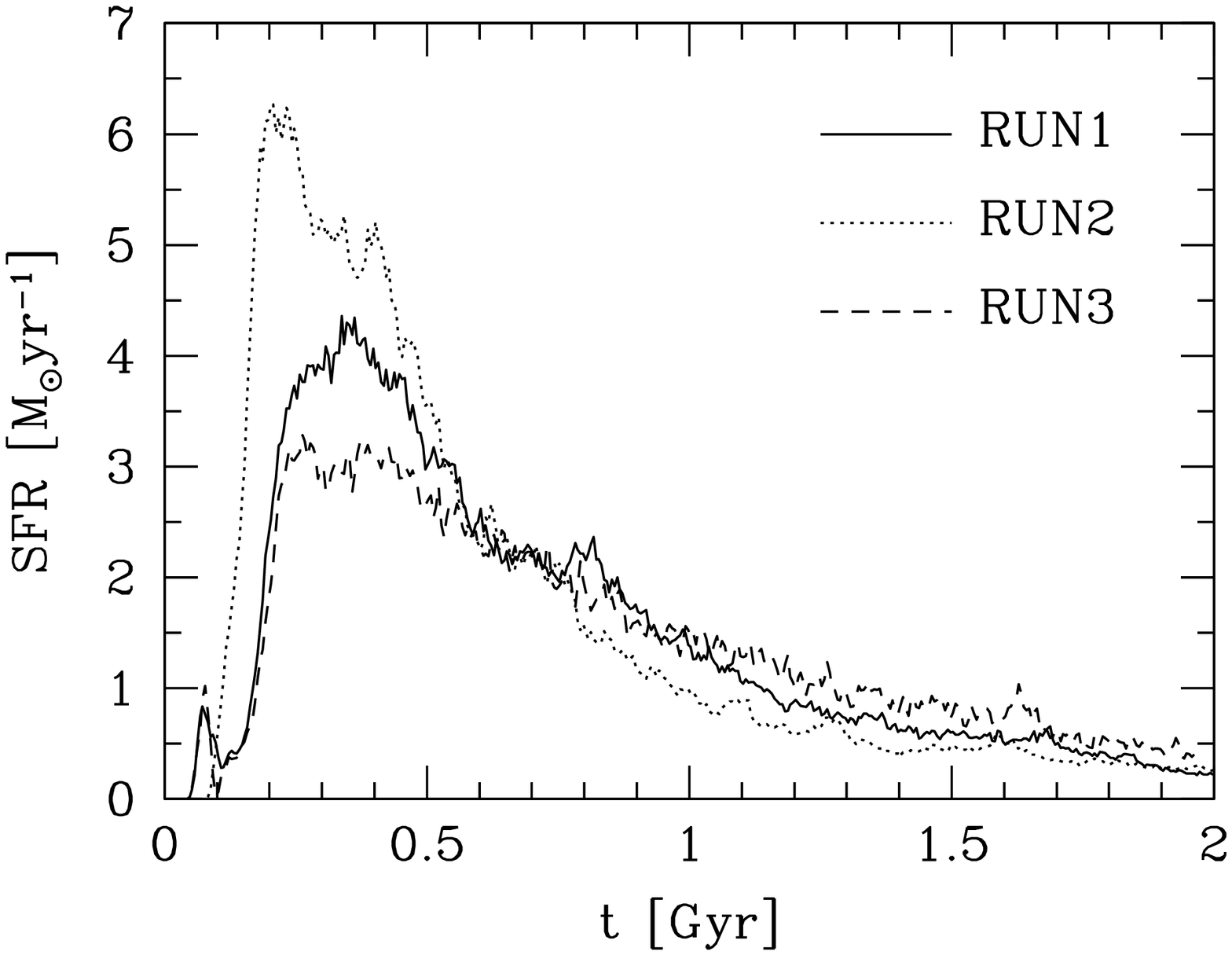,width=200pt} &
\psfig{file=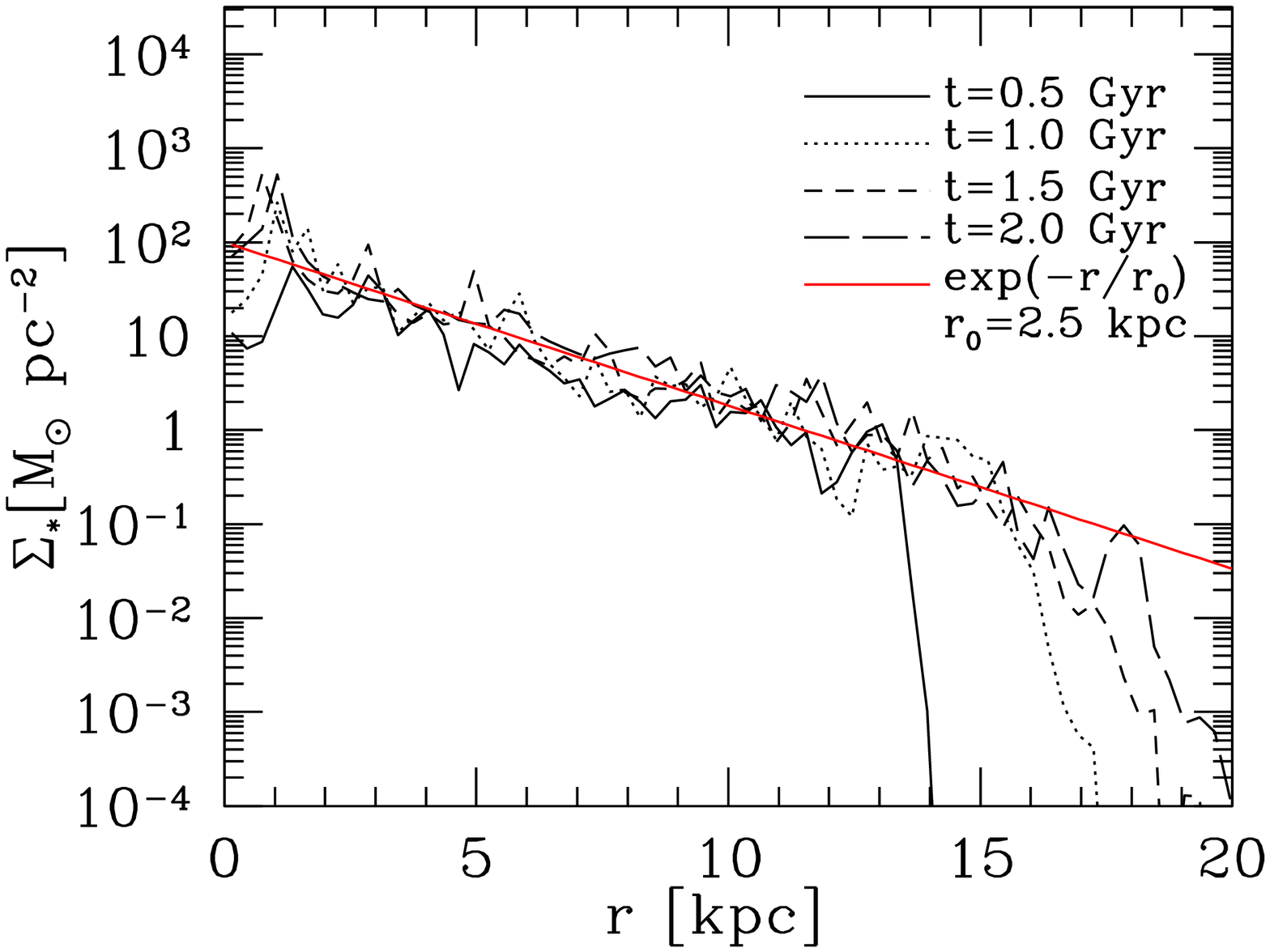,width=200pt} \\
\end{tabular}
\caption[]{(\emph{Left}) The star formation rate over time for RUN1, RUN2 and RUN3. The higher resolution in RUN2 allows for more star formation in the outer disc, while the feedback of RUN3 lowers the efficiency. At late times the SFRs show little difference. (\emph{Right}) Evolution of the stellar surface density. The density profiles are at all times well fitted with an exponential function. The red line is for $r_0=2.5\kpc$.} 
\label{fig:SFR}
\end{figure*}
The left panel in Fig.\,\ref{fig:SFR} shows the SFR of RUN1, RUN2 and RUN3. The main difference between the simulations is found in the most active star forming time, $t\sim 0.2-0.5\Gyr$, when the initial gravitational and thermal instabilities have formed dense clouds. The higher resolution in RUN2 allows for cloud formation in the less dense outer parts of the disc, leading to a higher SFR. The SNe feedback in RUN3 dampens star formation as explosions heat or disperse star forming clouds. However, after this star-bursting period the disc settles to a quiescent phase where all simulation approach a SFR $\sim 0.25 M_\odot{\rm yr}^{-1}$.

The stellar surface density is at all times well-fitted with an exponential function, $\Sigma_*(r)\sim \exp(-r/r_0)$ with $r_0\sim 2.5\kpc$, out to a truncation radius which grows with time, see right panel in Fig.\,\ref{fig:SFR}. This may not come as a surprise as the initial gaseous disc is set up to be an exponential. At the end of the simulation time there are $\sim1.3 \times 10^{6}$ stars in RUN1, $8.8\times10^{6}$ in RUN2 and $2.5\times10^{6}$ in RUN3.
\subsection{Composition and state}
\label{sect:composition}
\subsubsection{A multiphase ISM}
\label{sect:multiphase}
\begin{figure*}
\begin{tabular}{cc}
\psfig{file=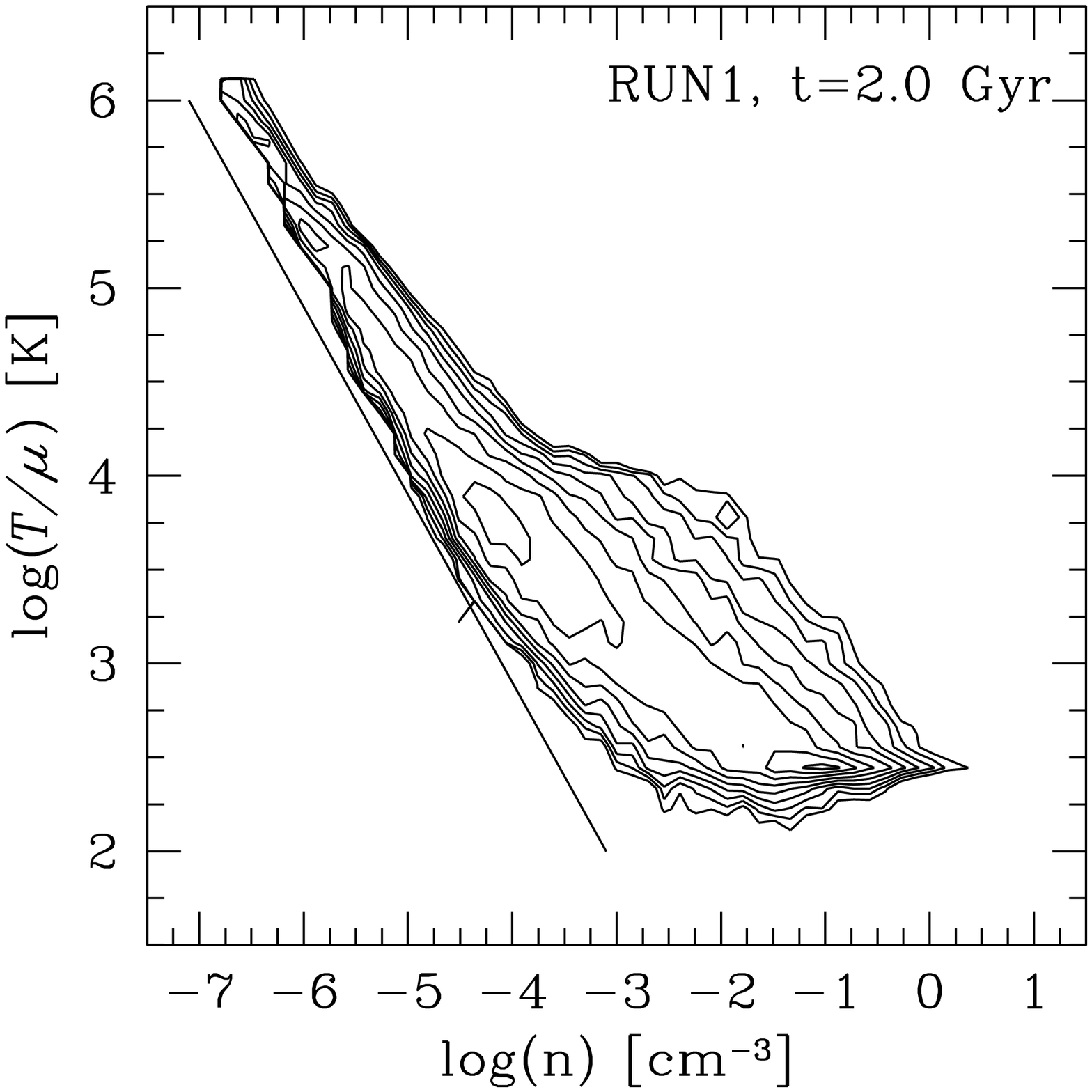,width=200pt} &
\psfig{file=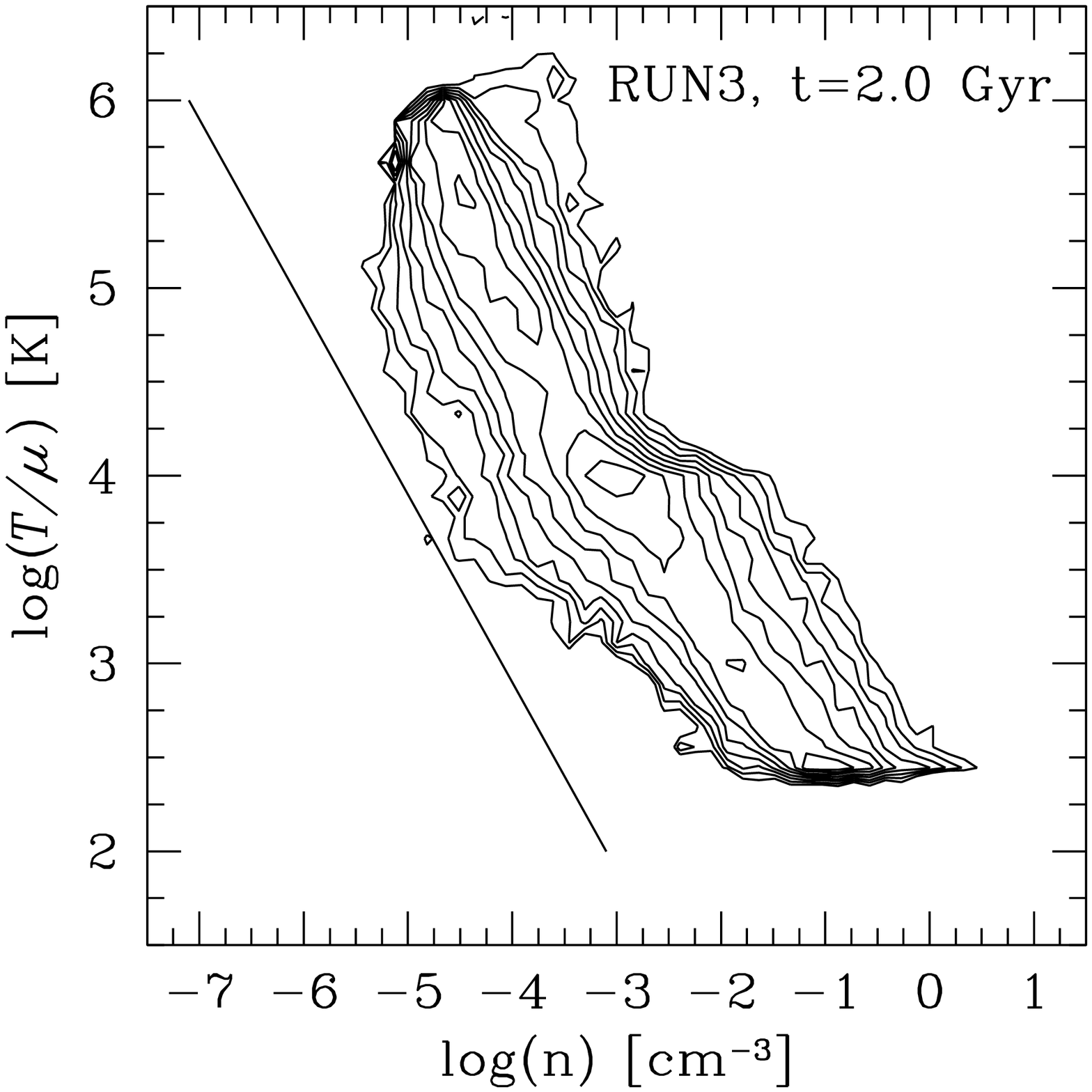,width=200pt} \\
\end{tabular}
\caption[]{Phase diagrams for RUN1 (\emph{left}) and RUN3 (\emph{right}) at $t=2.0\Gyr$. The solid straight line indicate the isobar.}
\label{fig:Trho}
\end{figure*}
\begin{figure*}
\begin{tabular}{cc}
\psfig{file=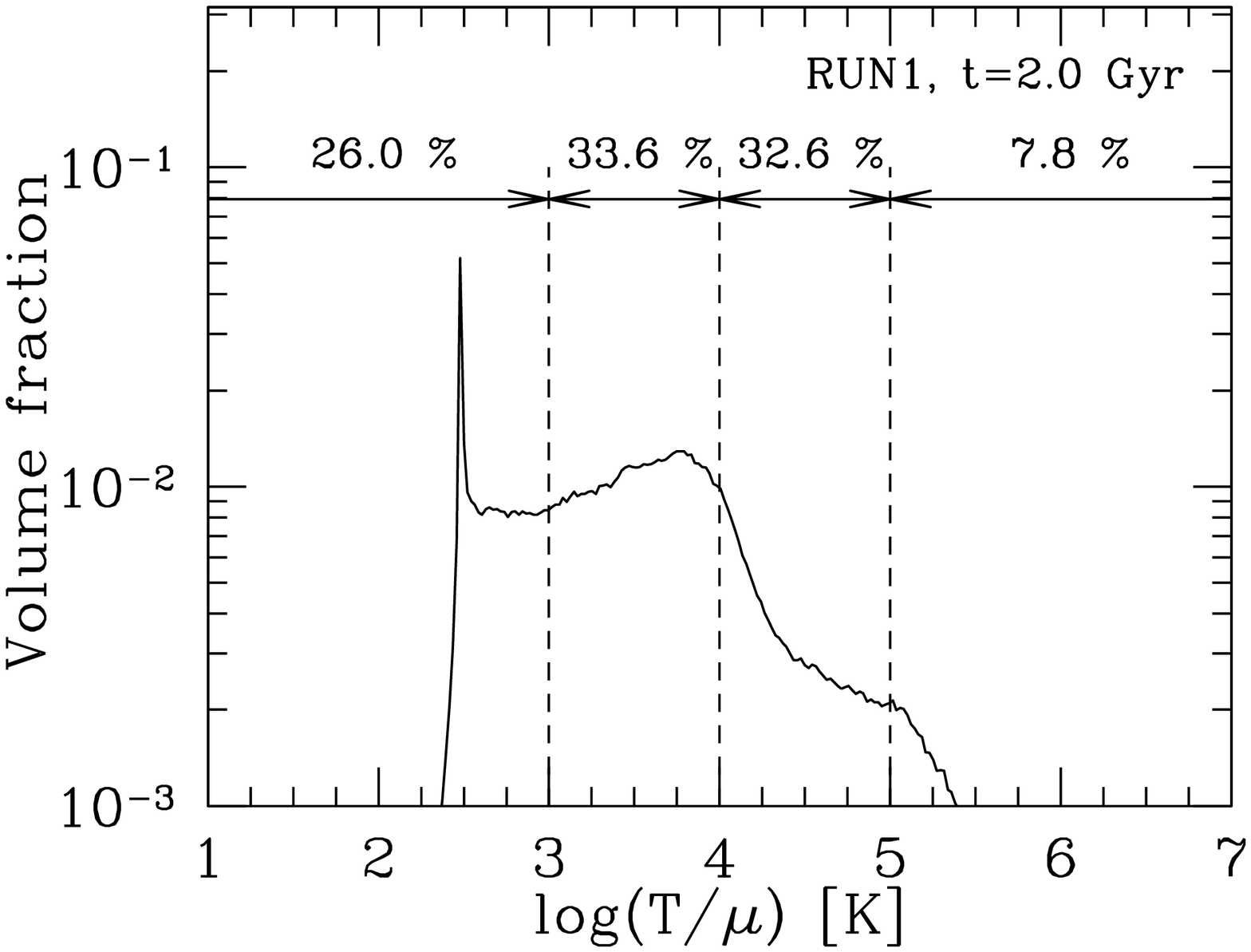,width=200pt} &
\psfig{file=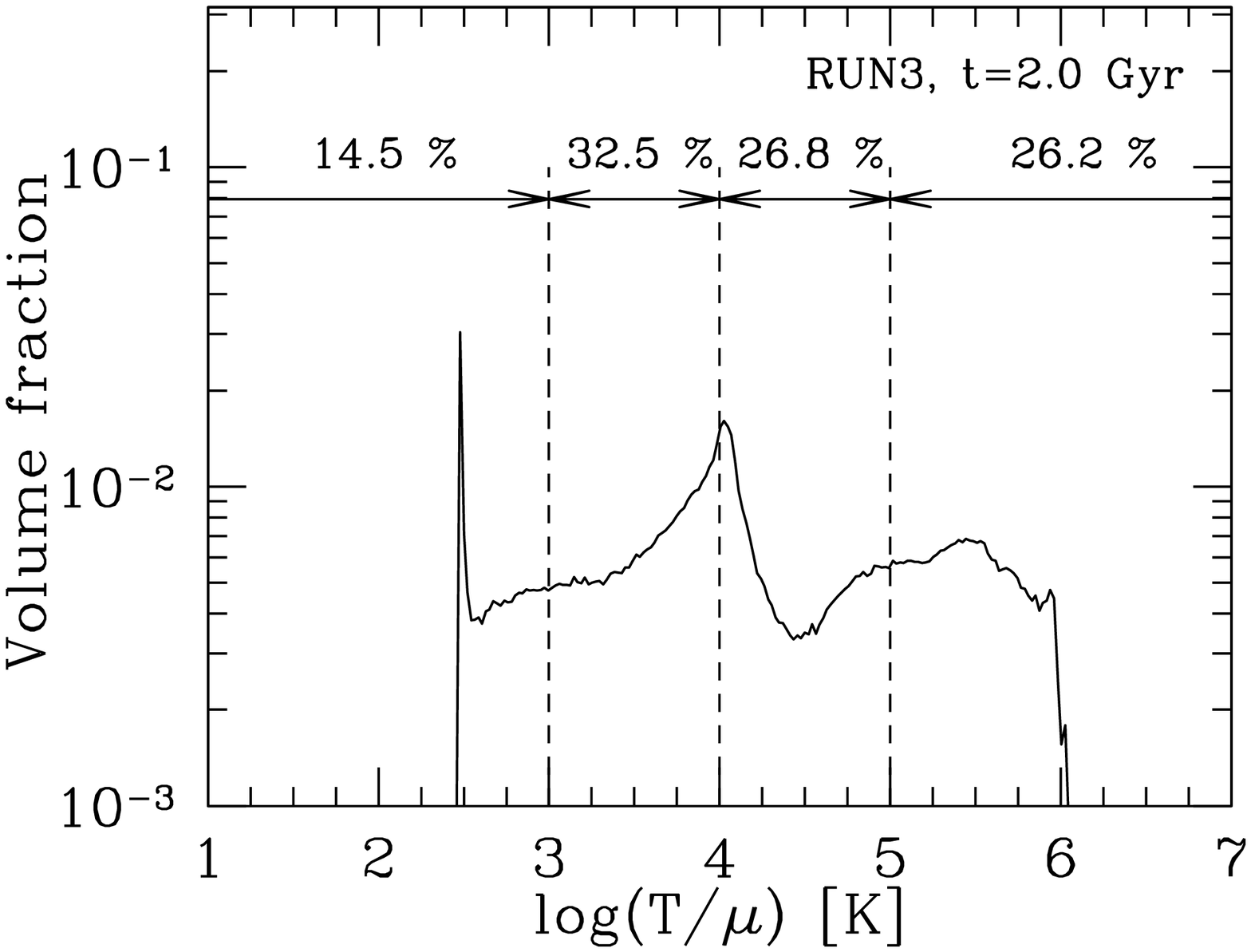,width=200pt} \\
\end{tabular}
\caption[]{Volume fraction for RUN1 (\emph{left}) and RUN3 (\emph{right}) at $t=2.0\Gyr$. While both models clearly show a cold and warm gas phase, the hot phase is only present in RUN3.}
\label{fig:Tvol}
\end{figure*}
Theoretical models of the ISM \citep[e.g.][]{mckeeostriker77} have a three phase structure consisting of a cold, warm and hot phase in pressure equilibrium where regulation is obtained through the balance of radiative cooling and supernovae heating. More updated models separate the phases further based on their ionization state. In this work we refer to the phases as cold ($T\lesssim10^3\,{\rm K}$), warm ($10^3\,{\rm K}\lesssim T \lesssim 10^5\,{\rm K}$) and hot ($10^5\,{\rm K}\gtrsim T$). A realistic ISM is very complicated which can be seen in the volume-weighted phase diagrams of RUN1 and RUN3 in Fig.\,\ref{fig:Trho}. Both runs show a wide range of temperatures and densities. RUN3 clearly displays distinct cold, warm and hot phases aligning to an isobaric strip, i.e. $P\sim\rho T\sim {\rm constant}$. Around this region we observe a large spread in both temperature and density. This analysis is approximately valid for RUN1 even though the warm and hot gas are smeared over less distinct phase regions and sits at slightly lower densities compared to RUN3. The cold phase is almost identical in the two models. The existence of a hot tenuous phase in RUN3 is due to supernovae heating, but why do we find hot gas in RUN1? As the initial circular velocity is set for the whole computational domain, the low density ambient gas at $T=10^4\,{\rm K}$ experiences a mild shock heating and settles into pressure equilibrium with the denser galactic disc. Some of the hot gas can also be found in the cloud-induced shocks in the disc.

Fig.\,\ref{fig:Tvol} shows the volume fraction occupied by gas at different temperatures. We also indicate the total percentage of gas in different temperature regions. The expected two-phase structure in RUN1 is evident. We find a clearly peaked cold phase with a transition into a warm phase in between $1\,000$ K and $10\,000$ K, peaking at $6\,000-7\,000$K which is the thermally unstable regime. The origin of the warm phase is shock-heating. RUN3 shows the same cold phase but the warm phase now strongly peaks at $10\,000$ K, just at the maximum peak of the cooling function. A hot gas phase is clearly present even though very little gas exists above $10^6\,{\rm K}$. As in RUN1, the warm phase dominates the gas volume.

It is desirable to approximately reproduce a mass distribution of molecular, cold atomic, warm atomic, warm ionized and hot ionized gas that agrees with observations (see e.g. \cite{ferriere01} for the Galactic inventory). However, even among the local group spirals there can be significant differences between the phase-distributions. For example, M31 has $\sim 40\,\%$ of its gas in cold HI, Milky Way $\sim25\,\%$ and M33 only $\sim15\,\%$ \citep{dickey93}. These differences could be due to the variation in baryon to dark matter fraction as we move down the Hubble sequence. As we show later, the formation of cold clouds is particularly sensitive to the gaseous disc mass. 
However, it is still instructive to compare our phase values of RUN1 and RUN3 at $t=1.5\Gyr$ to those of \cite{ferriere01} for the Milky Way. Roughly $50\,\%$ of the Milky Way gas is in molecular and cold atomic ($50-100\,\rm{K}$) clouds. Our simulations only allow for cooling down to 300 K and can hence not discriminate between the coldest gas phases.  By labeling all dense gas of $T<350\,\rm{K}$ as a joint cold cloud phase we find that $\sim55(47)\,\%$ of the total gas mass in RUN1(3) is cold. The warm neutral gas phase ($10^3\,\rm{K}<T<10^4\,\rm{K}$) has  $\sim11(16)\,\%$ while the total neutral mass fraction of the gas outside of clouds ($350\,\rm{K}<T<10^4\,\rm{K}$) is $39(45)\,\%$ respectively. The former value is lower than the Milky Way value ($\sim40\,\%$) which could be due to the fact that our initial conditions are more suitable for comparison with Sc galaxies. Also, including a homogenous UV background field in the simulations would heat the diffuse HI gas which seems to be the case in similar studies \citep[e.g.][]{bottema03}. Furthermore, the observed mass of warm phase in the Galaxy is derived from the observed HI velocity dispersion of $6-9\,\kms$ under the assumption of \emph{only} thermal broadening \citep{ferriere01}. A turbulent component can allow for the existence of colder gas yet retaining the velocity dispersion values. We will explore this notion further in Sect.\,\ref{sect:veldisp}.
\subsubsection{Disc stability}
\label{sect:stability}
\begin{figure}
\center
\psfig{file=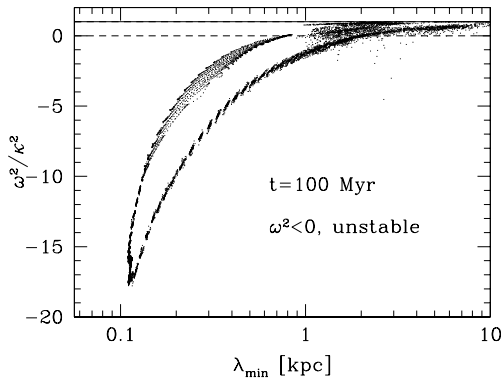,width=200pt} 
\caption[]{Most unstable wave lengths for the whole disc around time of fragmentation i.e. $t\sim 100\,$Myr. A large part of the disc is unstable at $\sim 100\pc - 2\kpc$, scales that will collapse to the initial cloud distribution.}
\label{fig:stabw2}
\end{figure}
\begin{figure}
\center
\psfig{file=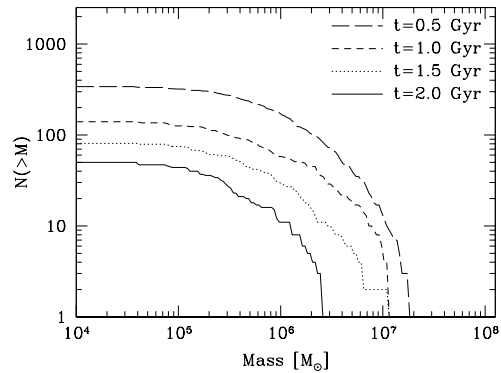,width=200pt} 
\caption[]{Cumulative mass spectrum of individual "molecular" clouds ($n>100\,{\rm cm}^{-3}$) in RUN3. } 
\label{fig:cloudspec}
\end{figure}
\begin{figure*}
\begin{tabular}{cc}
\psfig{file=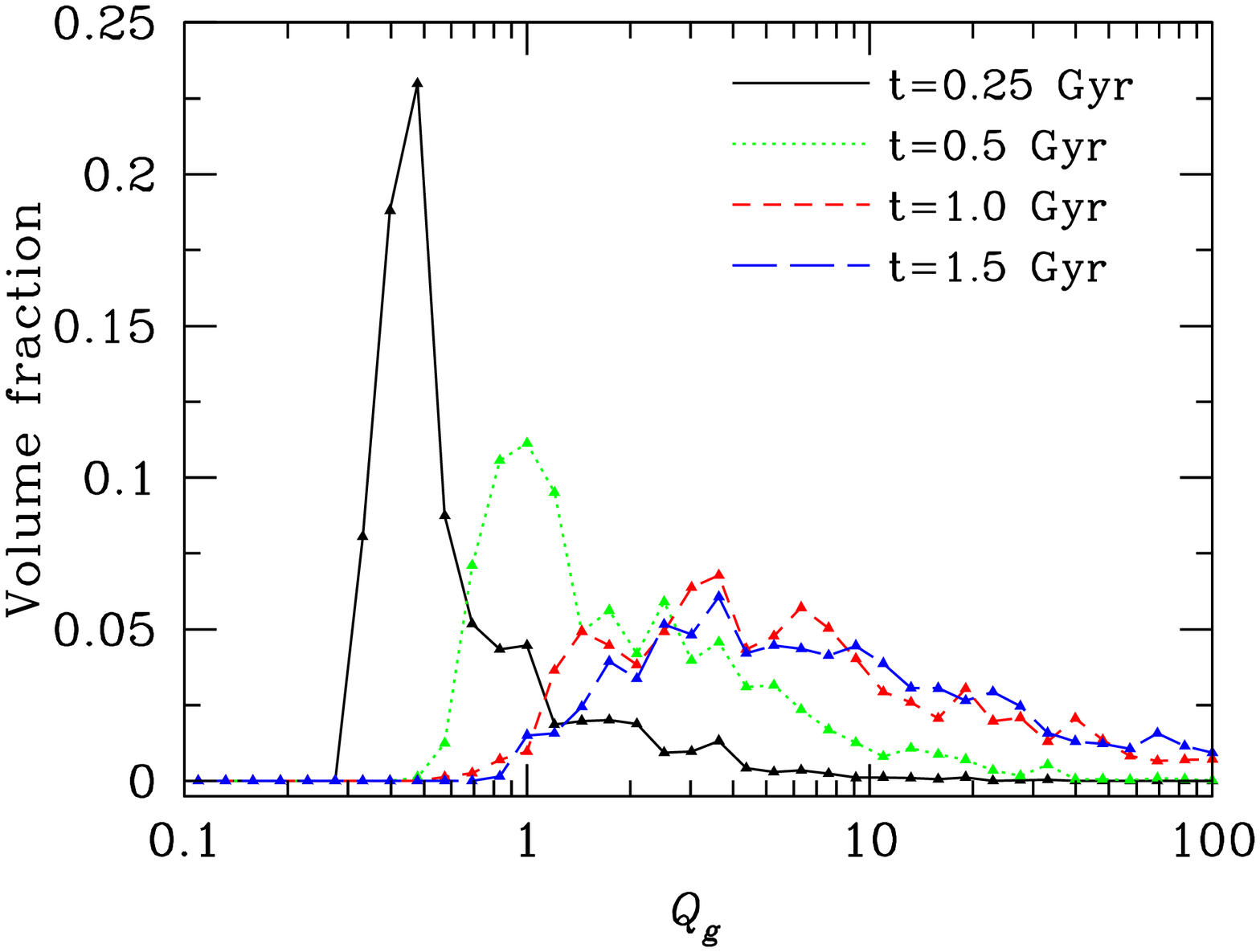,width=200pt}  &
\psfig{file=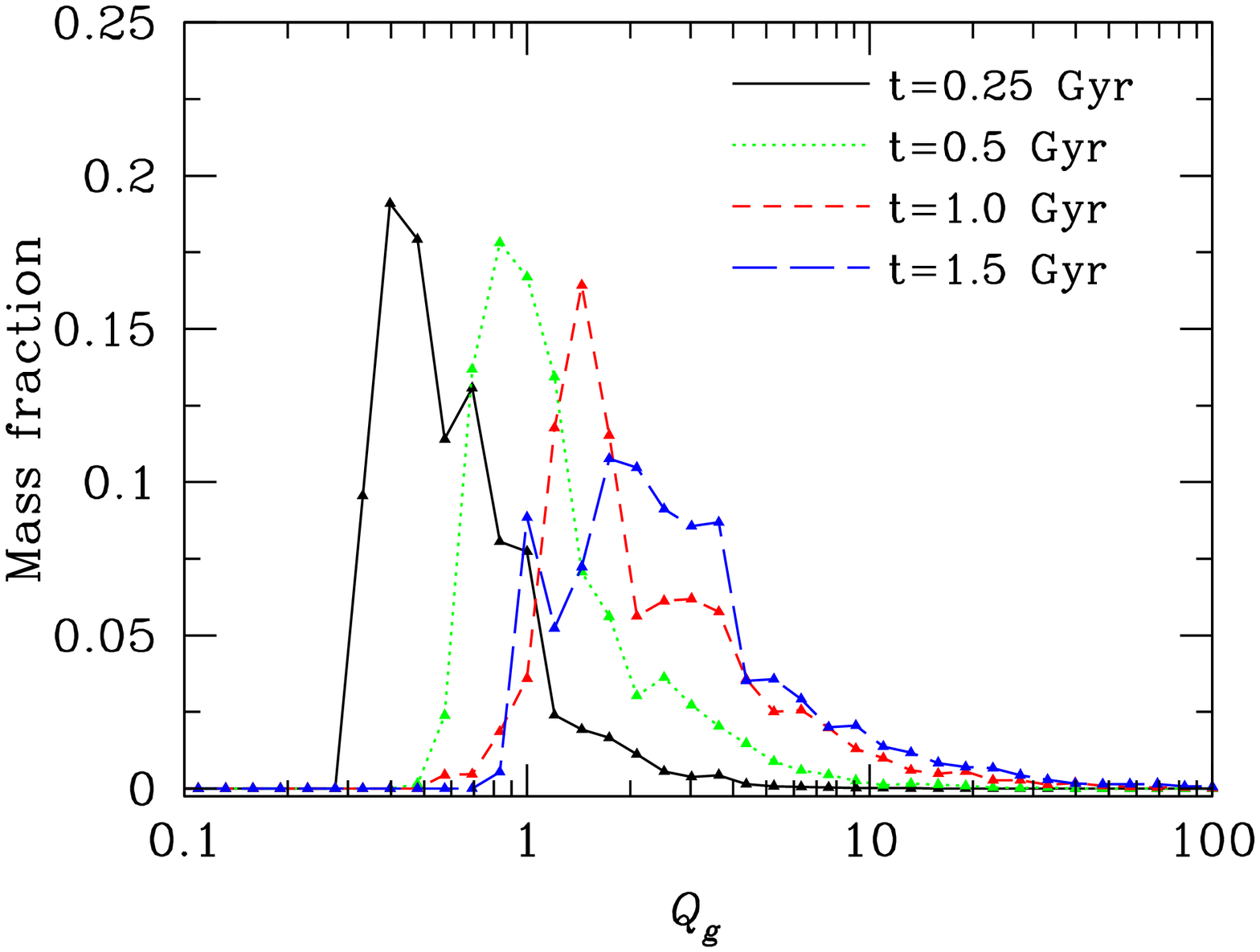,width=200pt}  \\
\end{tabular}
\caption[]{(\emph{Left}) Volume distribution of $Q_{\rm g}$ values for the whole disc in RUN1. The low $Q_{\rm g}$ values at early times indicates the formation and existence of clouds. Later times shows an equilibrium distribution that changes little with time. (\emph{Right}) Mass distribution of $Q_{\rm g}$ values. Most of the disc is distributed over $1<Q_{\rm g}<4$.}
\label{fig:qdisp}
\end{figure*}
To understand the relevance of gravitational instabilities in the simulated multiphase discs, we use the Toomre parameter \citep{toomre64} defined, for gas \citep{goldreichlyndenbell65a}, as
\begin{equation}
Q_{\rm g}=\frac{\kappa c_{\rm s}}{\pi G\Sigma_{\rm g}},
\end{equation}
where $c_{\rm s}$ is the sound speed of the gas. Since the gas has turbulent motions it is appropriate to use the effective
dispersion $\sigma^2_{\rm eff}=c^2_{\rm s}+\sigma^2_{\rm 1D}$, where $\sigma^2_{\rm 1D}$ is the average of the full three-dimensional velocity dispersion. The Toomre parameter is valid for local axisymmetric perturbations of two-dimensional discs, where $Q_{\rm g}<Q_{\rm c}=1$ implies instability. However, $Q_{\rm g}$ has been shown to characterize the response of discs to general gravitational instabilities. A finite disc thickness weakens the surface gravity and lowers the critical value where the disc undergoes instability. For example, \cite{goldreichlyndenbell65a} showed that $Q_{\rm c}=0.676$ for a single-component thick disc. In addition, the onset for \emph{non-axissymmetry} occurs at higher values of $Q_{\rm g}$, both for 2D ($Q_{\rm g}\sim1.7$) and 3D discs. Extended stability analysis taking thickness and multiple components (collisional and/or collisionless has developed by e.g. \cite{jogsolomon84,romeo92,rafikov01}. In this section we mainly focus on the more unstable gas component as it will be the main driver of turbulence compared to the stellar component which shows a higher degree of stability at all times ($Q_*>2$).  

We start by investigating the early time evolution when the initial cloud population forms. To do this we need to quantify the most unstable length scales. The dispersion relation for axisymmetric disturbances
\begin{equation}
\label{eq:disp}
\omega^2=\kappa^2-2\pi G\Sigma_{\rm g} k+\sigma_{\rm eff}^2k^2,
\end{equation}
where $\omega$ is the growth rate and $k$ is the wavenumber of the perturbation. Instability demands that $w^2<0$ and the most unstable mode is simply the minima of Eq.\,\ref{eq:disp}, i.e.
\begin{equation}
\lambda_{\rm min}=\frac{2\sigma_{\rm eff}^2}{G\Sigma_{\rm g}}.
\end{equation}
We calculate all components the dispersion relation, and $Q_{\rm g}$, on a polar grid with subregions $30\kpc/\Delta R=40$ and $2\pi/\Delta \theta=60$. Fig.\,\ref{fig:stabw2} shows the minimum of the dispersion relation and their related growth factors at $t=100\,$Myr. The bimodal distribution of points might seem odd but is merely a reflection of the initial $Q(r)$ (see Fig.\,\ref{fig:IC}), where the same value of $Q_{\rm g}$ exists at different radii (and different densities) and hence show the same $w^2$ at different $\lambda_{\rm min}$. We note that scales of $100\pc<\lambda_{\rm min}<2\kpc$ are in the unstable regime, where the smaller scales show larger negative values of $\omega^2$. It is reassuring that smaller scales remain stable due to the imposed Jeans capturing EOS discussed in Sect.\,\ref{sect:numconc}. These gravitational instabilities set the initial conditions for the clouds. The evolution of the cumulative mass spectrum, for "molecular" gas ($n>100\,{\rm cm}^{-3}$), in RUN1 is shown in Fig.\,\ref{fig:cloudspec}. We have calculated the mass spectrum by simply discerning individual pieces of high-density gas in the disc. This method is crude and occasionally overestimates the mass of clouds in the central parts of the discs where the gas density is high and crowding artificially identifies several clouds as one. Disregarding this, we note that the spectrum around $t\sim 1.5\Gyr$ occupies similar values as that of local group spirals \citep{blitz07}, where the most massive clouds are $\sim 10^7\,M_\odot$. The small mass truncation is due to limited resolution. As the simulations lack important small scale physics, e.g. MHD, radiative transfer, cosmic rays etc., the cloud population is long-lived and only reflects a true ISM \emph{in a statistical sense}.  This notion should not be a problem for the source of turbulent velocity dispersions that, as shown in Sect.\,\ref{sect:turb}, is due to large scale gravitational drags that most probably is independent of the small scale gas state close to or inside of the cloud complexes.

We now turn to the subsequent evolution. Fig.\,\ref{fig:qdisp} shows the time evolution of the distribution of $Q_{\rm g}$ values for the whole disc. The left panel shows the volume fraction that different values of $Q_{\rm g}$ occupy while the right panel treats the mass fraction.
This figure illustrates the complexity of the simulated discs and why azimuthally averaged $Q_{\rm g}(r)$ can be misleading. Initially, the disc shows a low spread of $Q_{\rm g}$ around a value of a few. As the disc cools down and undergoes gravitational instability (after $t\sim0.1 \Gyr$) this simple picture changes. At $t=0.25\,\Gyr$, the disc has undergone fragmentation and the distribution is confined to $0.2<Q_{\rm g}<1$. The peak of the distribution, and the dispersion, gets larger with time. Part of this owes to star formation that acts to lower $\Sigma_{\rm g}$. For $t>1.0\,\Gyr$ the disc evolves into what appears to be an equilibrium state, spanning a large range in $Q_{\rm g}$-values ($0.5\lesssim Q_{\rm g}\lesssim 10^2$). This co-existence of $Q_{\rm g}$-value in a patchy galactic disc is in agreement with the analysis of \cite{wada02} for their two-dimensional models. The mass fraction distribution follows a similar evolution, approximately reaching an equilibrium state after $t>1.0\,\Gyr$. However, a significant part of the $Q_{\rm g}$ distribution at late times now populates the unstable or marginally stable values. At $1.5\,\Gyr$, most of the disc is distributed around $1\lesssim Q_{\rm g}\lesssim 4$. Regardless of the exact distribution, the dominating existence (by mass) of unstable and marginally stable regions of the disc is of great importance for generating a global gravitoturbulent state which we will return to in Sect.\,\ref{sect:driver}. Without the onset of gravitational instabilities, the gas would approximately stay on circular orbits. 

At late times the mass in the disc is dominated by the stellar component. To get an understanding of its influence on stability one can use an approximate stability parameter \citep{bertinromeo88,romeo94} which in standard regimes is of the form  
\begin{equation}
Q\approx Q_*\left(1-2\frac{\Sigma_{\rm g}}{\Sigma_*}\right).
\label{eq:Qcombo}
\end{equation}
At late times, the stellar component is in the range $1.5\lesssim Q_* \lesssim 5$ in the star forming region which together with the values shown in Fig.\,\ref{fig:qdisp} assures us that the multicomponent disc will never be completely stable, at least locally.

\subsection{The turbulent ISM}
\label{sect:turb}
Having characterized the multiphase ISM in terms of phases and stability, we can now properly address the main topic of this work, the HI velocity dispersions.
\subsubsection{Velocity dispersions}
\label{sect:veldisp}
\begin{figure*}
\begin{tabular}{cc}
\psfig{file=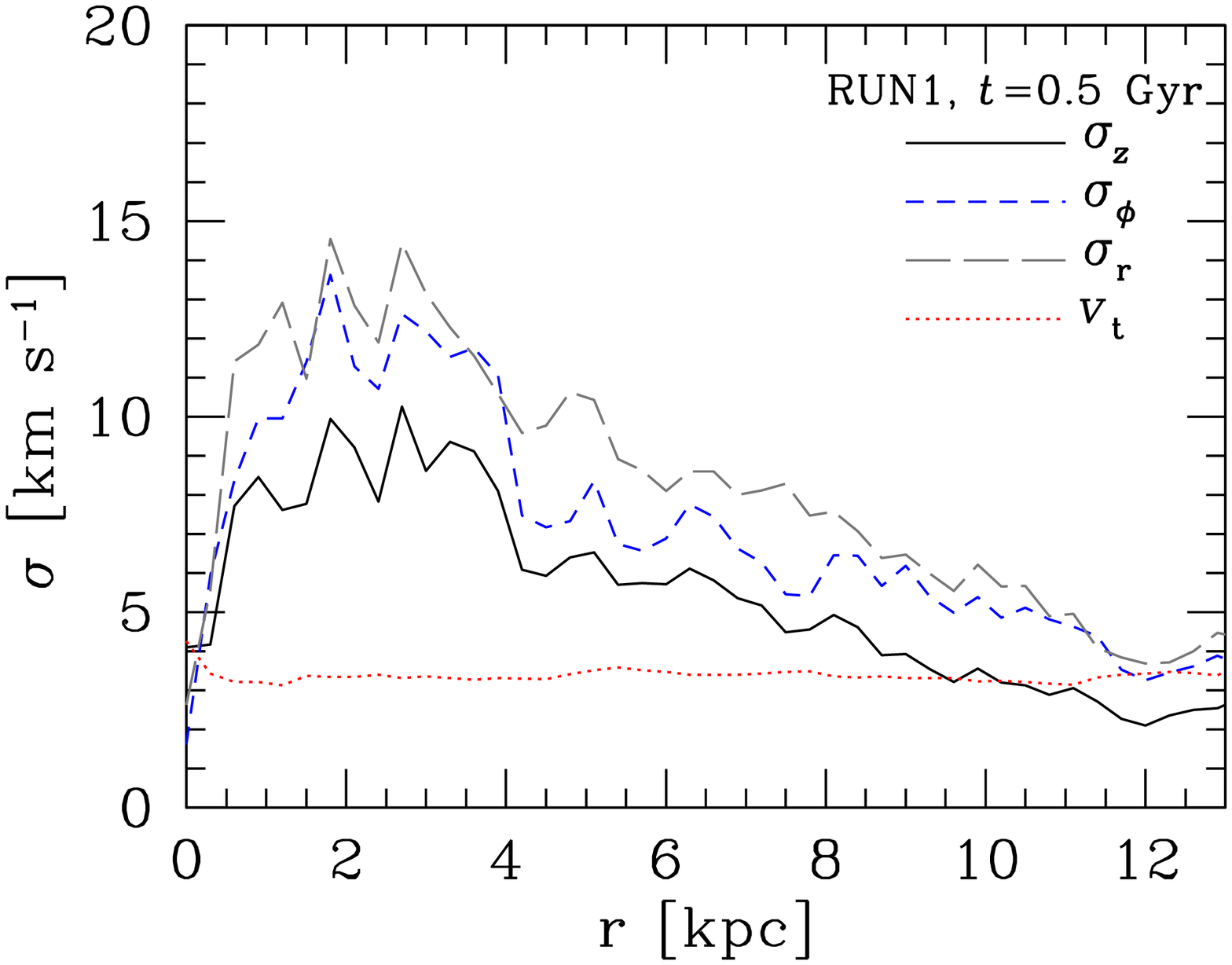,width=200pt} &
\psfig{file=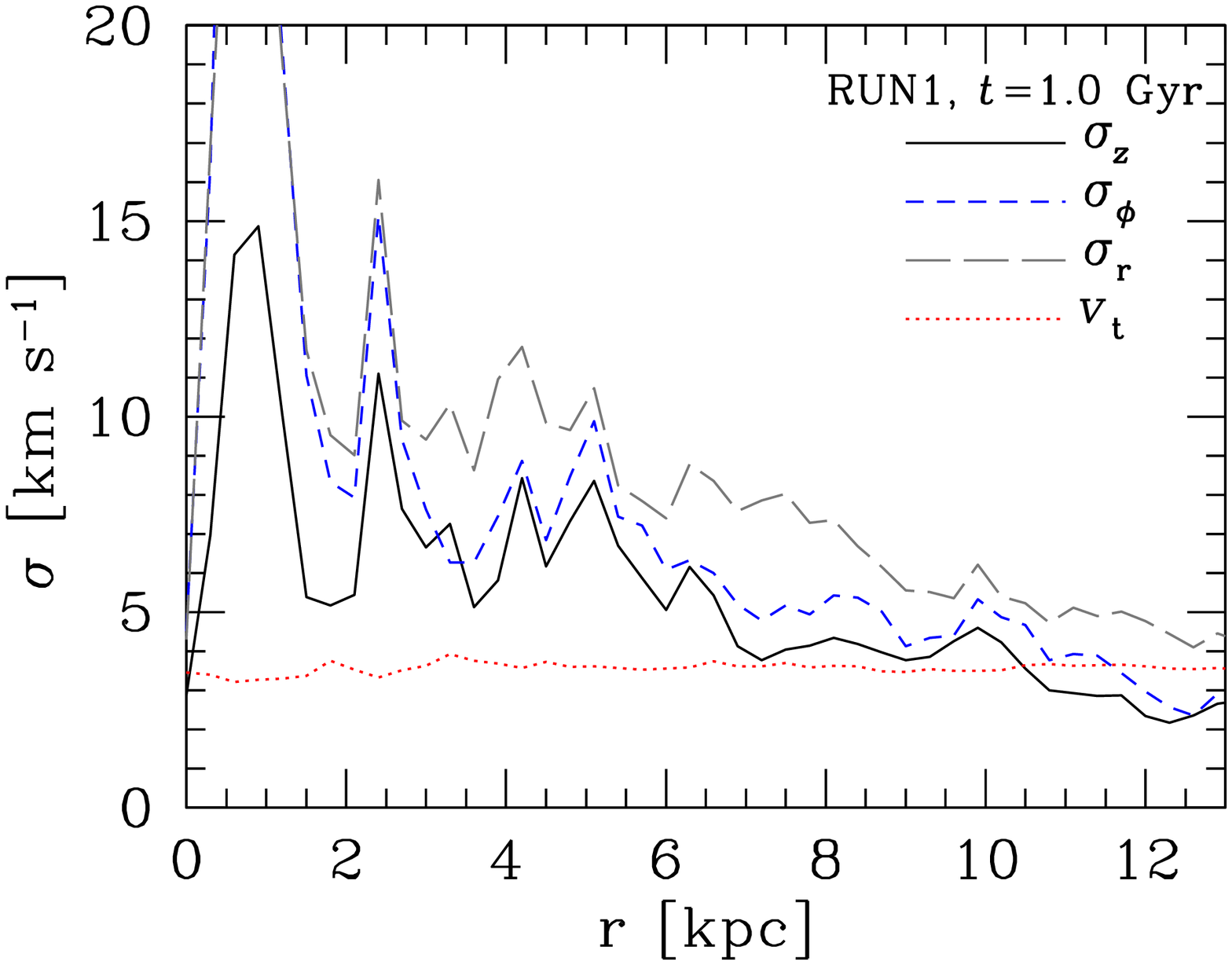,width=200pt} \\
\psfig{file=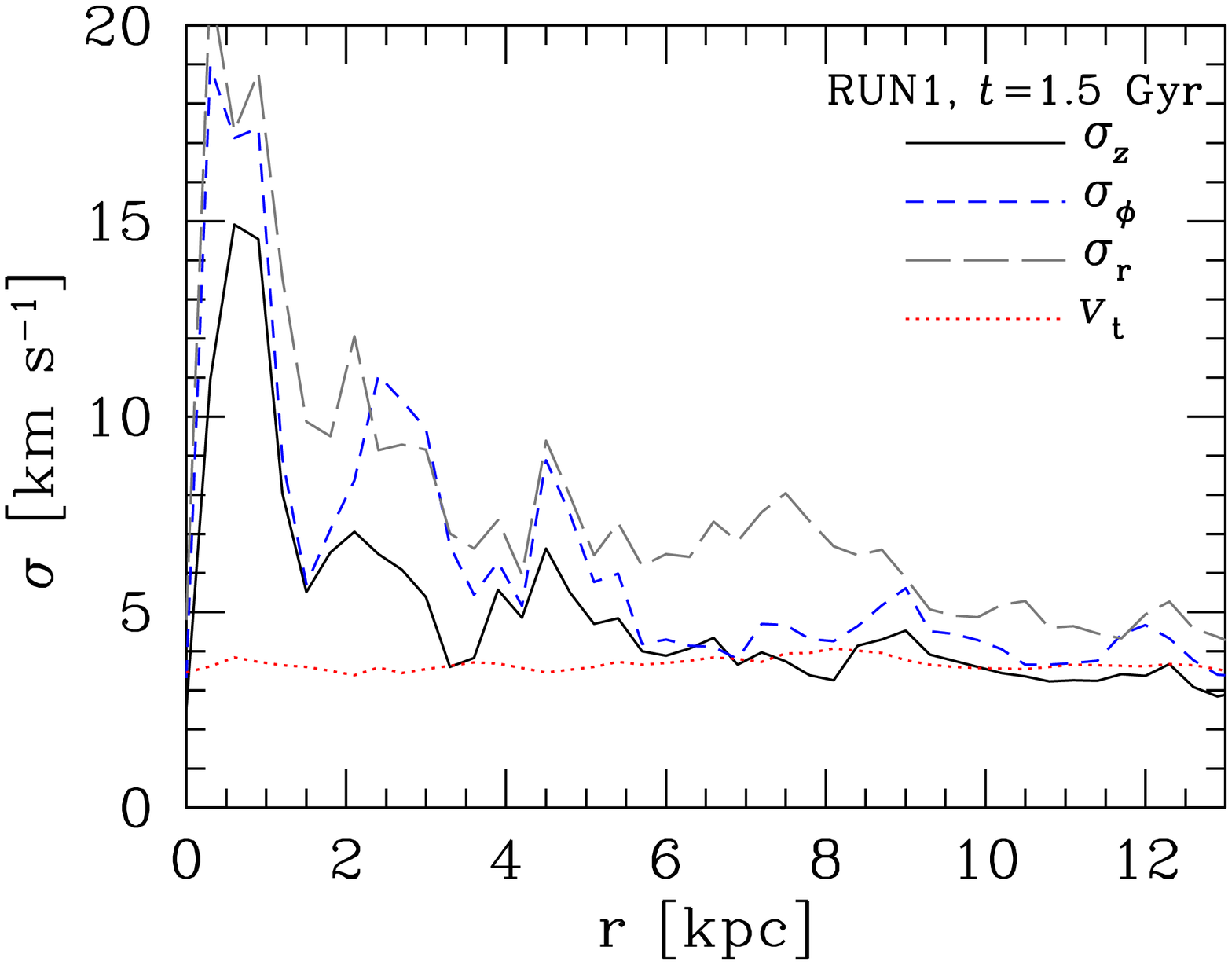,width=200pt} &
\psfig{file=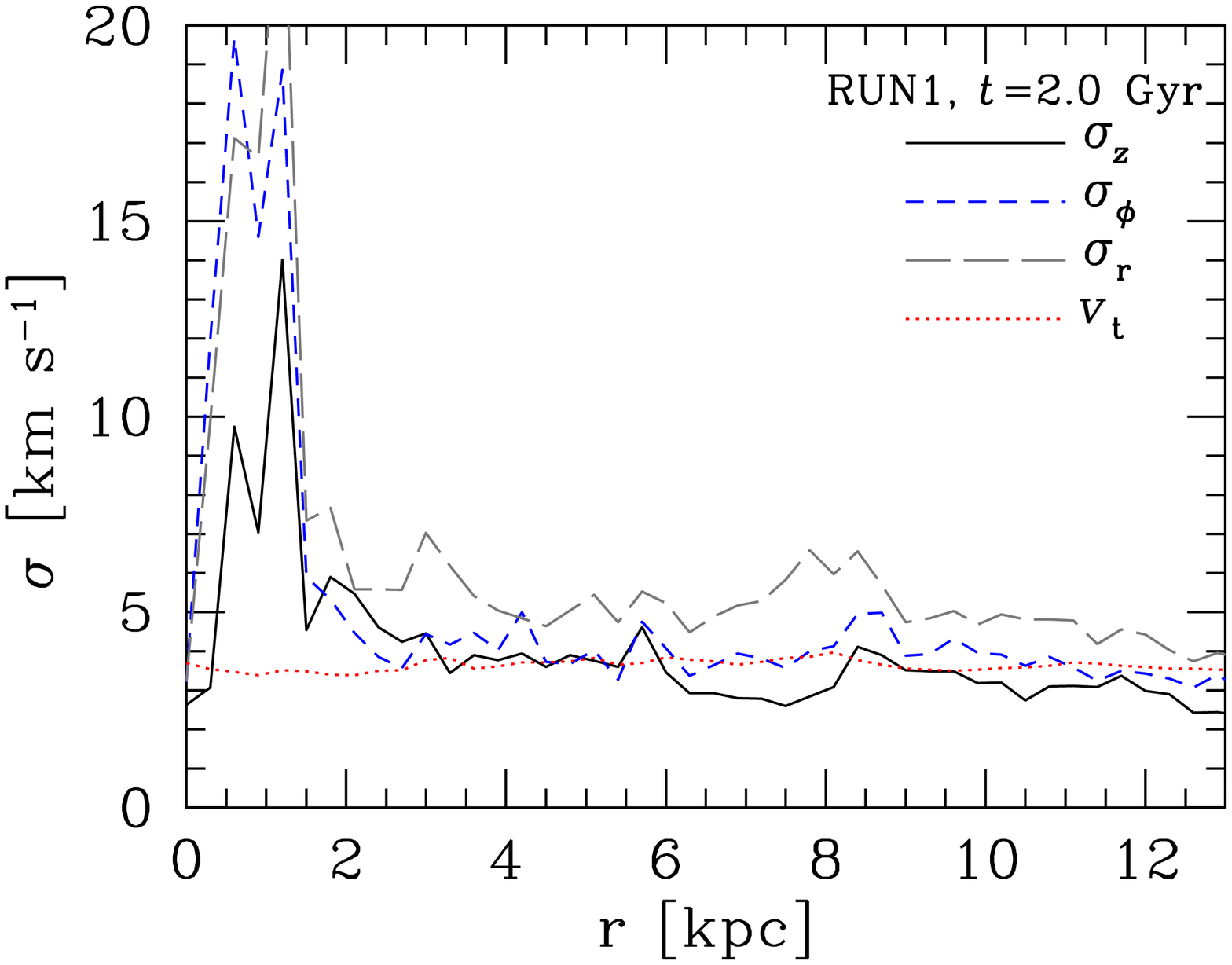,width=200pt} \\
\end{tabular}
\caption[]{Velocity dispersions of the vertical ($\sigma_z$), angular ($\sigma_\phi$), radial ($\sigma_r$) and thermal component ($v_{\rm t}$) of the HI gas in RUN1 at $t=0.5, 1.0, 1.5$
 and $2.0\Gyr$.}
\label{fig:veldisp}
\end{figure*}
\begin{figure}
\center
\psfig{file=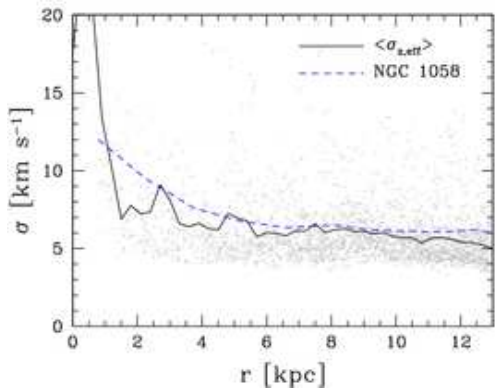,width=200pt} 
\caption[]{Observed velocity dispersion (blue dashed line) of the face on galaxy NGC 1058
 compared to the effective vertical dispersion (black solid line) of
 RUN1 at $t=1.5\Gyr$. Grey dots indicate locally measured
 dispersions, see text.} 
\label{fig:veldispobs}
\end{figure}
The observational tradition is to model HI profiles using one or multiple Gaussians where the flux is a function of the velocity $v$ as
\begin{equation}
f(v)=\frac{1}{\sigma\sqrt{2}}\exp\left[-\left(\frac{1}{2\sigma^2}\right)(v-v_0)^2 \right]
\end{equation}
where $v_0$ is associated with the peak flux and $\sigma$ is the actual velocity dispersion. Broadening of spectral lines is mainly due to thermal and Doppler/turbulent effects. We will discuss the thermal effects in terms of the thermal velocity $v_{\rm t}=\sqrt{RT/\mu}$ (i.e. the isothermal sound speed of the gas), where $R$ is the gas constant and $\mu$ is the molecular weight. Random bulk motion of the gas is quantified in terms of its turbulent velocity dispersion $\sigma_{\rm t}$. We calculate the net observable dispersion by adding the turbulent and thermal contribution in quadrature, i.e. $\sigma_{\rm eff}^2=v_{\rm t}^2+\sigma_t^2$. 

In the following analysis, we characterize only the gas that would be observed as HI and not the dense clouds that will consist of mainly molecular gas. We therefore use the criteria $\rho < 10\,{\rm cm}^{-3}$ (star formation threshold) and $800\,{\rm K}< T < 10\,000\,{\rm K}$. This choice is suitable as it is more likely to exist outside of the denser spiral arms as well as in the outer regions of the disc, which is where observations lack an explanation. The velocity dispersion is calculated by randomly sampling the galactic discs using synthetic observational patches ($5\,000$ patches were used for the data described here) of size $\sim 700\pc$. We choose this size as this is the stated scale below which bulk motions are expected to be responsible for the observed dispersions \citep{petric07}. In each patch we calculate both the mass weighted turbulent velocity dispersions $\sigma_{\rm t}$ and the mass weighted mean thermal velocity. Weighting by mass is well motivated as HI emission is strongly correlated with the local density.

The panels in Fig.\,\ref{fig:veldisp} shows a time evolution of the radial behaviour of the velocity dispersion for RUN1, where $\sigma_z$ is the vertical dispersion component, $\sigma_r$ the radial and $\sigma_{\phi}$ the angular. $\sigma_z$ show typical values of $\sim 15\kms$ in the center and declines to $\sim 3-5\kms$ at large radii. The velocity dispersion is clearly anisotropic as $\sigma_r>\sigma_\phi>\sigma_z$ at all times. It is interesting that the ratio of the dispersions roughly follow the epicyclic predictions for a collision-less system, i.e. $\sigma_r=2\Omega\sigma_\phi/\kappa$. A similar result was found by \cite{bottema03}. The planar dispersion  $\sigma_{xy}$, i.e. the RMS value of the radial and angular dispersions, is a factor of $\sim 2$ larger than the vertical dispersion at all times and radii. The thermal component of the gas lies in the range $3-5 \kms$ in agreement with a warm gas component ($T\sim1000-2000\,$K). The planar velocity dispersion is supersonic or transsonic at all times and radii while the vertical dispersion is generally transsonic, turning sub-sonic at large radii. This means that the thermal component becomes as important as the turbulent at large radii for the total observable velocity dispersion. By considering a minimal observable ($\sigma_{\rm eff}$) for the $z$-component, we clearly find an agreement with the observed HI dispersions values described in Sect.\,\ref{sect:intro} (i.e. $\sigma_{\rm eff}\sim12-15\,\kms$ in the inner parts declining to $\sim4-6\,\kms$ in the outer). Any inclination would boost these values due to the $\sigma$-anisotropy. The same analysis has been performed on the higher resolution simulation RUN2 with no significant difference in the results.  

We now directly compare our simulations to the HI data of the spiral galaxy NGC 1058 \citep{dickey90,petric07}. NGC 1058 is a suitable object for comparisons, as it is comparable in size, surface density and peak rotational velocity (derived to be $\sim 150 \kms$) to our simulated disc. The galaxy also has a low star formation rate, SFR\,$\sim 3.5\times 10^{-2}\,M_\odot\,\rm{yr}^{-1}$ \citep{ferguson98}, which places it in the flat part of Fig.\,\ref{fig:dib2006}. Furthermore, by being an almost perfectly face-on galaxy (inclination of $4-11\,^{\circ}$), we can disentangle the vertical component from the planar. In Fig.\,\ref{fig:veldispobs} we compare $\sigma_{z,{\rm eff}}$ of RUN1 at $t=1.5\Gyr$ with the observational data of NGC 1058. Our simulation not only reproduces the magnitude of the velocity dispersion but also the declining radial shape.
\begin{figure*}
\center
\begin{tabular}{cc}
\psfig{file=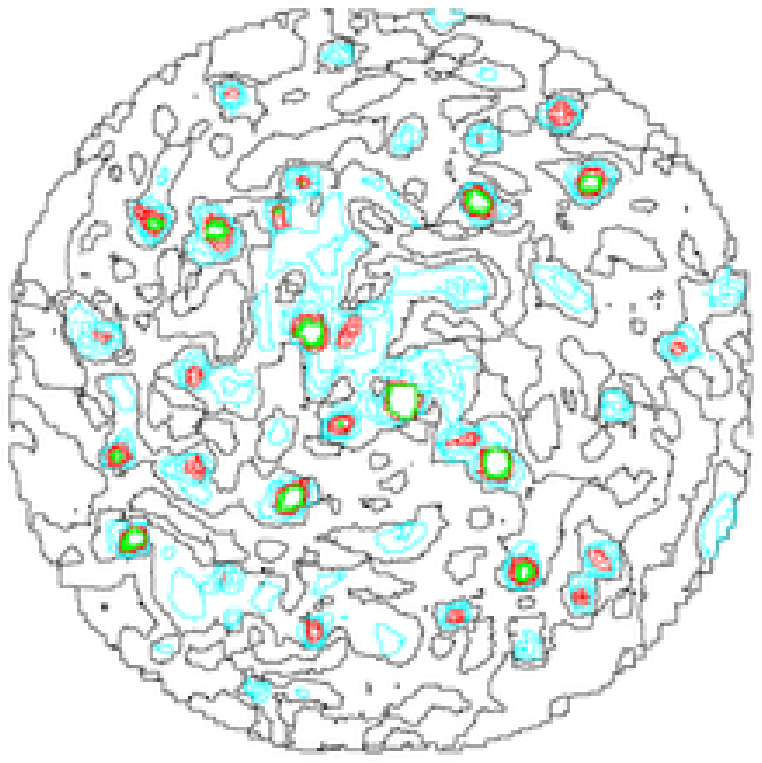,width=200pt} &
\psfig{file=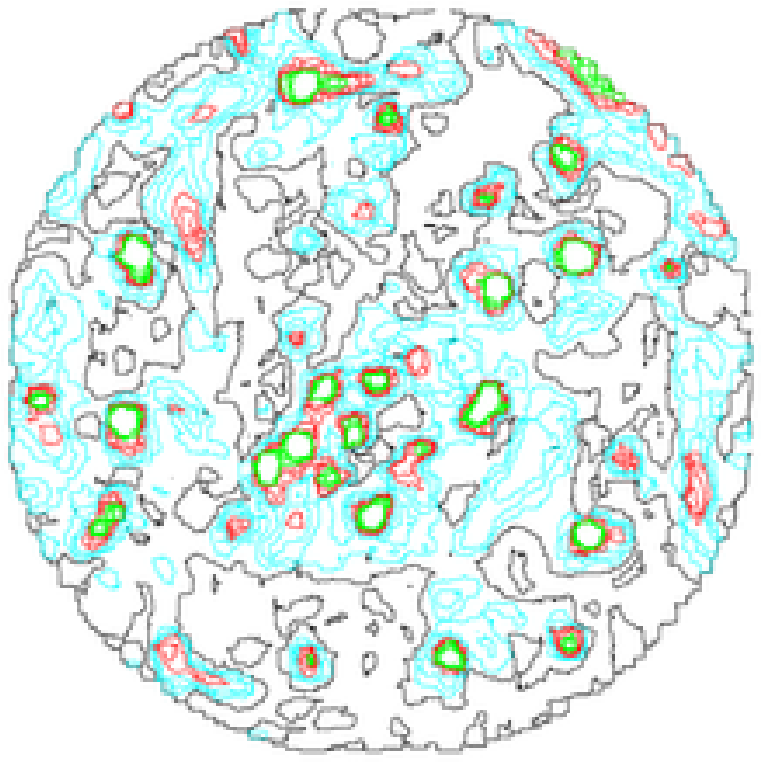,width=200pt} \\
\psfig{file=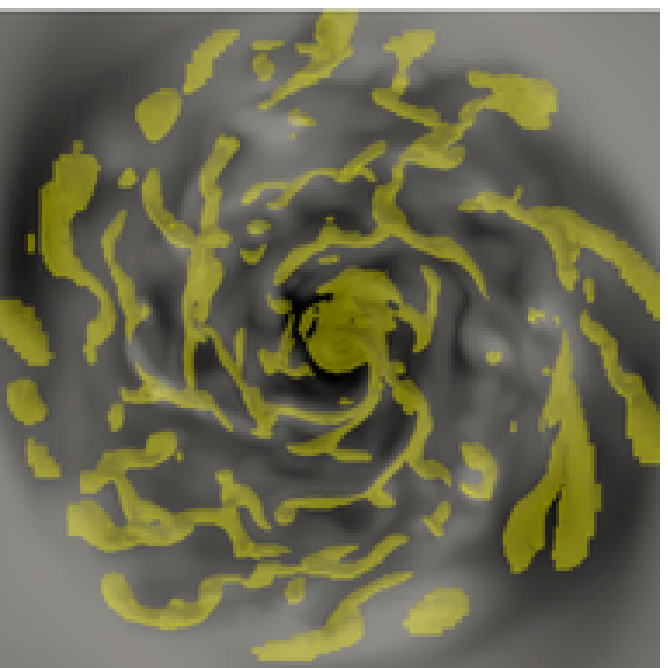,width=180pt} &
\psfig{file=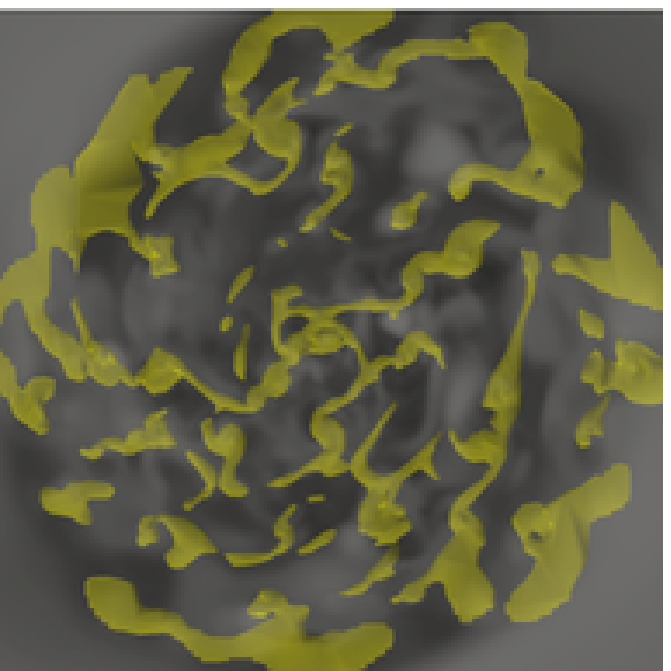,width=180pt} \\
\end{tabular}
\caption[]{Distribution of vertical velocity dispersions in RUN1 (\emph{top left}) and RUN3 (\emph{top right}) at $t=2.0\Gyr$ calculated for HI only (see text). The plotted region is $30\kpc$ across. The contour levels are in $\kms$ in steps of $1\kms$. Black is used between 3 and 5, cyan for 6 to 8, red for 9 to 11 and green for 12 to 14 $\kms$. The bottom panels show contours of the gas with $n>0.2\,{\rm cm}^{-3}$ of the same regions but slightly zoomed out ($40\kpc$ across).}
\label{fig:contoursigma}
\end{figure*}

The spatial distribution of the vertical velocity dispersions in NGC 1058 is very patchy with several peaks of $\sigma>10\kms$ (see Fig. 5 of \cite{petric07}). This observation is reproduced by our simulations, as shown in Fig.\,\ref{fig:contoursigma} where we plot the contours of $\sigma_{{\rm eff},z}$ in RUN1 and RUN3 at $t=2.0\Gyr$ as well as the corresponding density fields. The high density gas is distributed in a flocculant spiral structure, reminiscent of the HI observation of M33 \citep{deulhulst87,engargiola03}. Analyzing the simulation at a late epoch is preferred as the mass spectrum of dense clouds has evolved to a rather realistic state, see Fig.\,\ref{fig:cloudspec}. The strongest peaks ($\sigma_{{\rm eff},z}>10\kms$) are associated with dense clouds while mildly turbulent regions (cyan levels at $\sigma_{{\rm eff},z}\sim 6-8\kms$) often exist in inter-cloud/arm regions of strong shear. Regions of large velocity dispersion related to clouds also extend several $\kpc$ away from their radius of influence. We note that RUN3 even at late times displays a few $\kms$ larger velocity dispersions in diffuse regions, probably due the more prominent warm gas phase.

\subsubsection{What is the driver of ISM turbulence?}
\label{sect:driver}
\begin{figure*}
\begin{tabular}{ccc}
\psfig{file=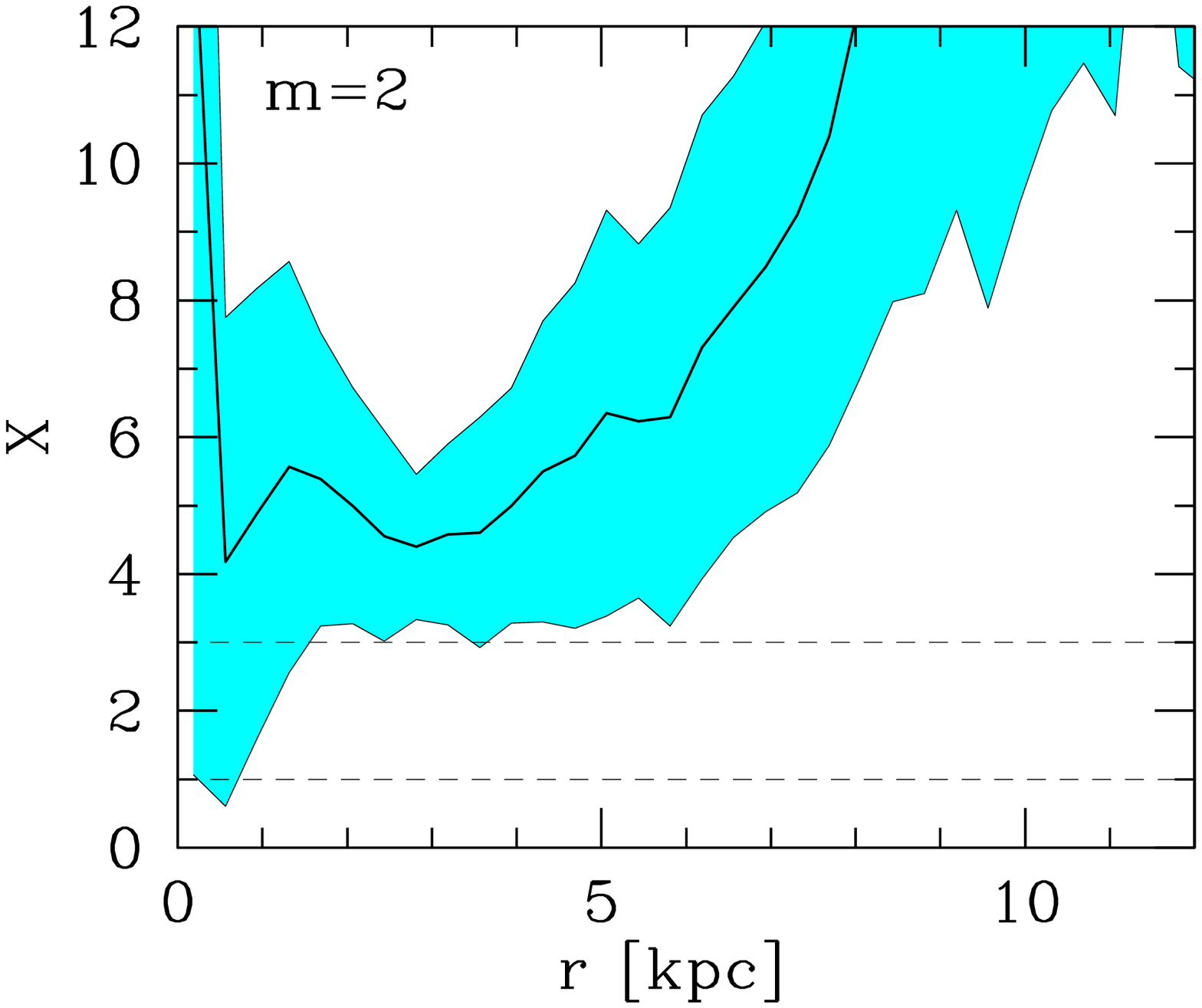,width=161pt} &
\psfig{file=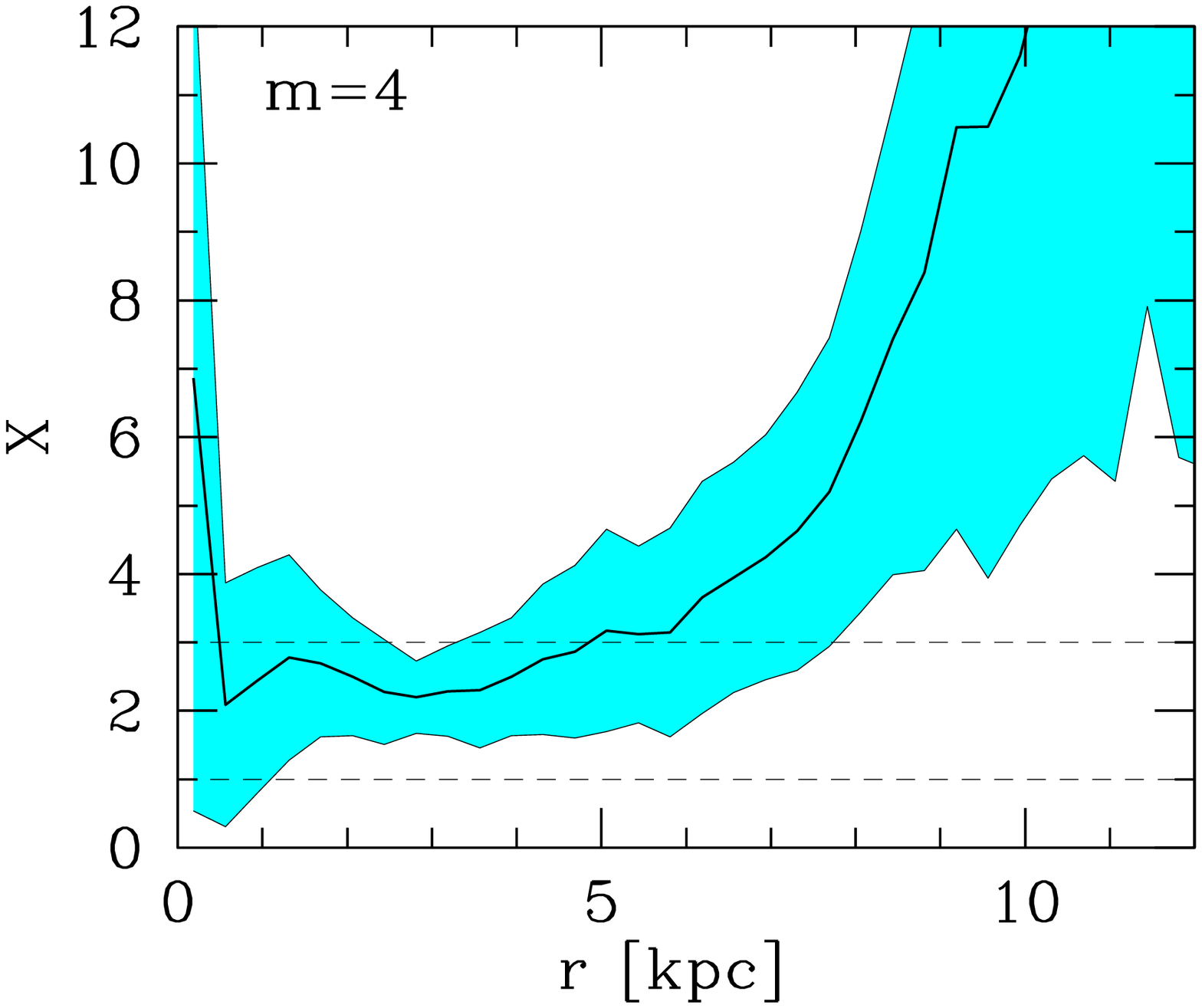,width=161pt} &
\psfig{file=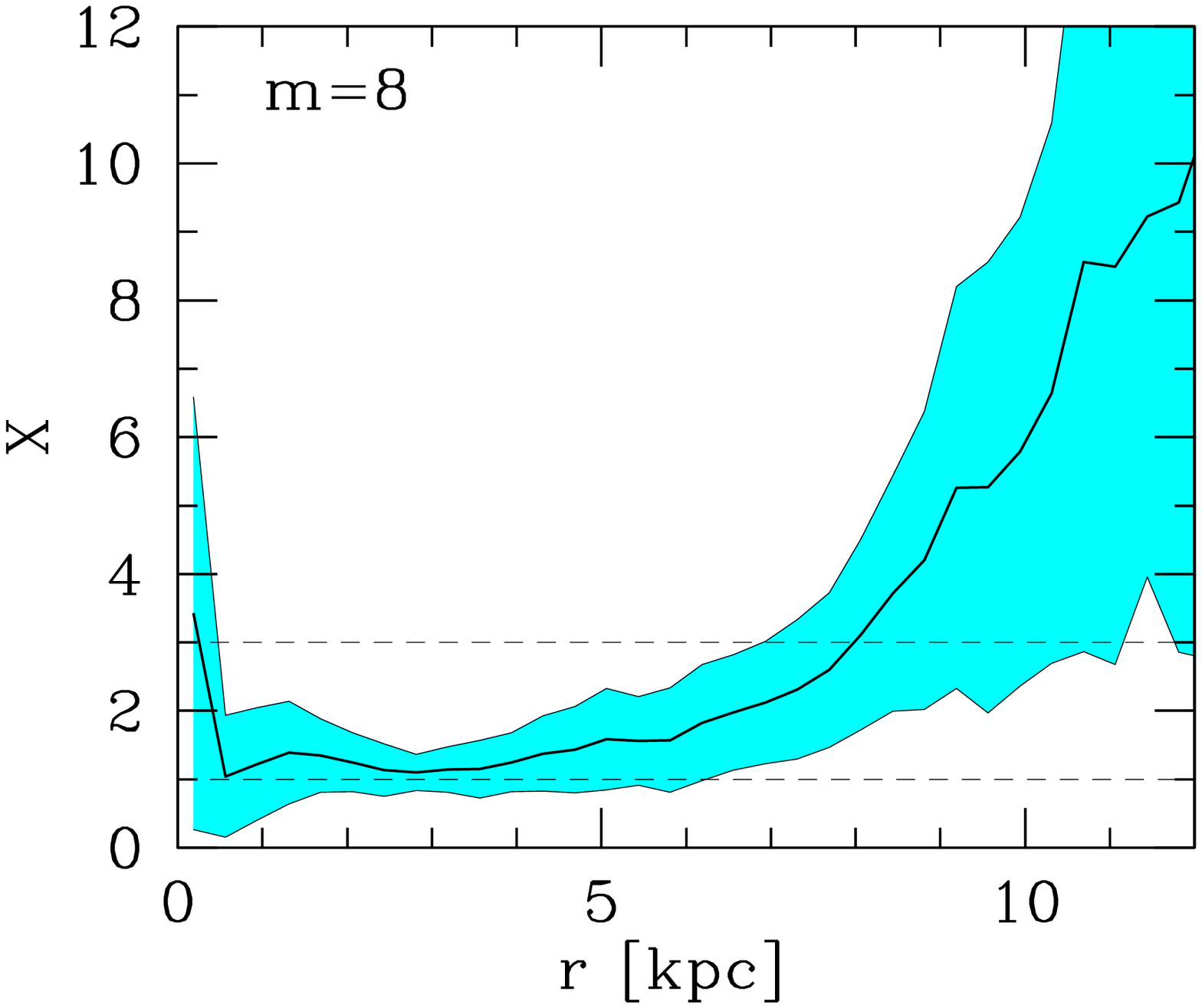,width=161pt} \\
\end{tabular}
\caption[]{Radial dependance of the $X$-parameter for $m=2,4$ and 8. Higher order modes ($m\gtrsim4$) are in the range $1<X<3$ where swing amplification is efficient.}
\label{fig:X}
\end{figure*}

The drivers of the turbulent component of the velocity dispersion are gravity and shear. In Sect.\,\ref{sect:stability} we found that the galactic discs have, by mass, a wide spectrum of $Q_{\rm g}$ values where a significant part sits at local marginal stability for a finite thickness disc. The 2D shearing box simulations by \cite{kimostriker07} showed that a marginally stable gas discs at $Q_{\rm g}\sim 1.2$ can generate velocity dispersions of the order of the local sound speed, decreasing for larger $Q_{\rm g}$-values (see their Fig. 12). The origin of turbulence was here attributed to swing amplification. Note that the 3D structure of our discs necessitates values $\sim 25\,\%$ lower for an equivalent stability. A full turbulent outcome of more unstable discs ($Q_{\rm g}<1.2$) was not studied as the velocity field would then only be a response to very strong density inhomogeneities. Local shearing boxes are useful for understanding the mechanism that drives turbulence at \emph{specific values} of $Q_{\rm g}$. As our simulations show a wide range of $Q_{\rm g}$ values we have the combined spectrum of swing amplified turbulence across the whole disc for gas that locally behaves in accordance with the simulations of \cite{kimostriker07} for $Q_{\rm g}>1.2$. In a statistical sense there will always be regions with a $Q_{\rm g}$-value low enough to tap large velocity dispersions from the swing mechanism which is confirmed in Fig.\,\ref{fig:contoursigma} where intermediate values of $\sigma_{{\rm eff},z}$ is associated with waves. To quantify this it is useful to use the $X$-parameter \citep{toomre81} defined as 
\begin{equation}
X=\frac{k_{\rm crit}R}{m}
\end{equation}
where $k_{\rm crit}=\kappa^2/2\pi G \Sigma$, $\Sigma=\Sigma_{\rm g}+\Sigma_*$ and $m$ is the number of arms. It has been shown by \cite{jog92} that swing-amplification is very effective in the gas component in multicomponent discs. Even when both the gas and stars separately are stable ($Q_{\rm g}=Q_*=2$), their gravitational coupling can amplify waves in he gas for values of $X$ not much larger than unity. In one component, marginal stability and $1<X<3$ can be considered sufficient to assure amplification \citep{toomre81}. Fig.\,\ref{fig:X} shows the radial dependance of the $X$-parameter for $m=2,4$ and 8 at $t=2.0\,\Gyr$. We see that amplification is efficient for $m\geq 4$ which confirms the high-order flocculant spiral structure in Fig.\,\ref{fig:contoursigma}. 

\begin{figure*}
\center
\psfig{file=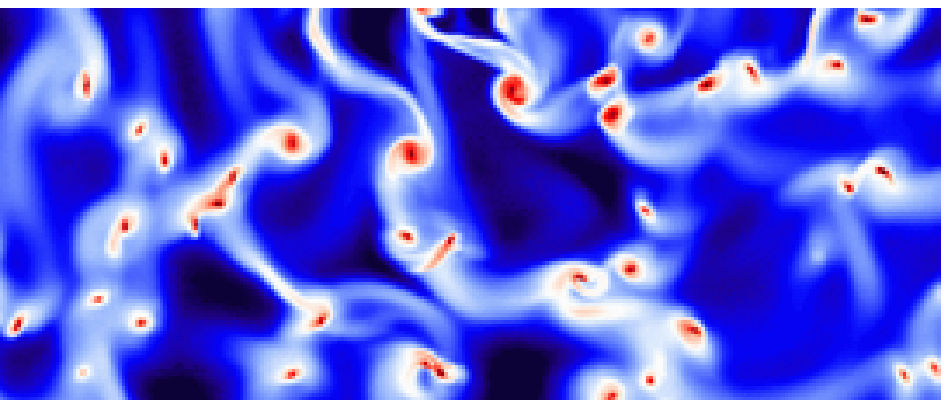,width=350pt} 
\caption[]{Density plot of a $24\times 10\kpc^2$ region centered over $\{x,y,z\}=\{0,-5.0,0\}\kpc$ in RUN1 at $t=1.0\Gyr$. The colour map is here chosen to enhance the visual appearance of the clouds and filaments. The galactic rotation is here clock-wise. Filamentary structures is always associated with the clouds. Also, all clouds excite waves, many of them leading which will swing into trailing ones.}
\label{fig:waves}
\end{figure*}

For small values of $Q_{\rm g}$, where the gas locally has undergone full non-linear gravitational instability, the situation is different. The cold phase dominates the gas mass, even at early times and is therefore locally the most important gravitational source. Direct cloud merging and tidal interactions stirs the inter-cloud medium both radially and vertically. Apart from stirring the gas, the clouds also dissipate energy thermally in shocks which regulates the warm phase of the ISM, forming the $\sim 4-5\kms$ thermal components of $\sigma_{\rm eff}$. We observe cloud formation, merging, scattering and reformation during the whole simulation time. Formation in a shearing environment causes dense structures that are not tightly bound to the actual clouds to stretch into waves and filaments. This triggering of wave-like perturbations, and its associated irregular velocity field in the ISM, is a key role of the clouds which was realized already by \cite{juliantoomre66}. As for the marginally stable gas surrounding the gas, the leading waves swing and amplify, inducing gravitational torques in the gas and hence increasing the local velocity dispersion \citep[e.g.][]{kimostriker02,kimostriker07}. Fig.\,\ref{fig:waves} shows a typical patch of the disc in RUN1 at $t=1\Gyr$, confirming this notion. We note that this processes is analogous to the energy extraction from background shear at a rate $T_{R\phi}d\Omega/d\ln R$, where the $T_{R\phi}$ stress tensor includes the contribution from Reynolds and Newtonian stresses, to induce local velocity dispersion as outlined by \cite{sellwoodbalbus99}.
\begin{figure}
\center
\psfig{file=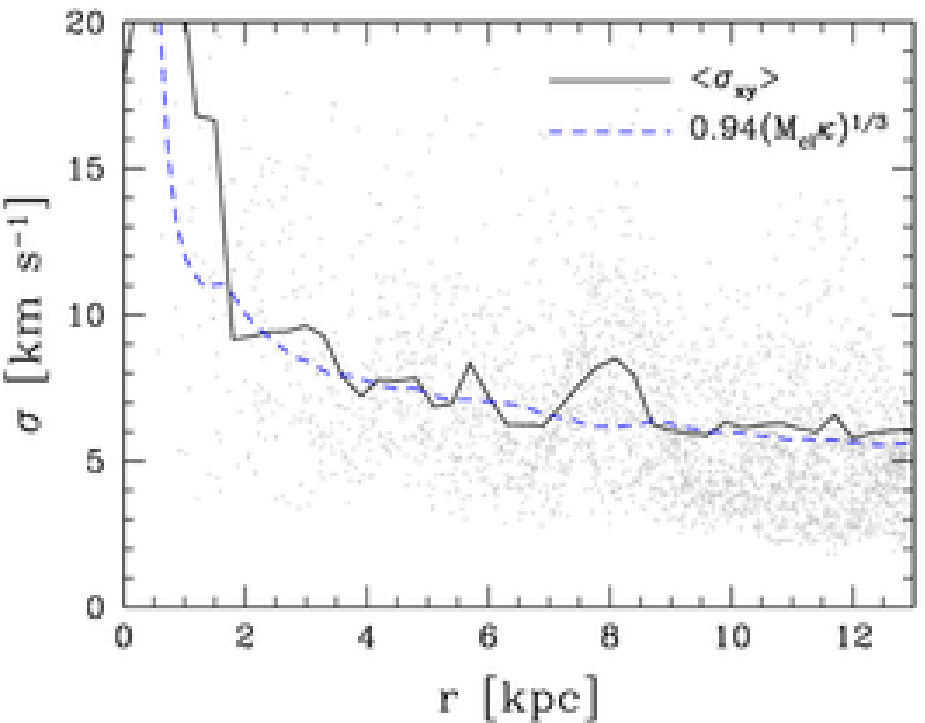,width=200pt} 
\caption[]{Planar velocity dispersion of RUN1 at $t=2.0\Gyr$ (black
 solid line) compared to the relation derived by Gammie et al. (1991)
 (blue dashed line) for cloud scattering. The epicyclic frequency is
 obtained from the simulation and the cloud mass is chosen to be
 $3.5\times 10^6 M_{\odot}$.} 
\label{fig:veldispplane}
\end{figure}

\emph{But how can we quantify the impact of the cloud motions?} Let us assume that the motions of the cloud ensemble are representative of the turbulent ISM. The swing amplifier might play a very fundamental role for turbulence but as the clouds effectively trace the large scale waves and constitute the majority of the mass, the assumption that turbulence is associated with cloud motions is a fairly good approximation. Cloud-cloud interaction can be modelled as gravitational scattering and has been studied analytically by e.g. \cite{jogostriker88} and \cite{gammie91}. The semi-analytical perturbation theory model of \cite{gammie91} predicts a planar velocity dispersion 
\begin{equation}
\label{eq:sigmagammie}
\sigma_{xy}\approx 0.94(GM_{\rm cl}\kappa)^{1/3},
\end{equation}
where $M_{\rm cl}$ is a typical mass of a cloud. This relation is derived for a two-dimensional, two body encounter on radially separated orbits in a shearing disc. However, as these clouds are the main local perturbers by mass we can assume that any diffuse HI gas will approximately be dictated by the cloud ensemble velocities. In Fig.\,\ref{fig:veldispplane} we plot $\sigma_{xy}$ for RUN1 at $t=2.0\Gyr$ against Eq.\,\ref{eq:sigmagammie} using a cloud mass $M_{\rm{cl}}\approx 3.5\times 10^6 M_\odot$, and $\kappa(r)$ of the gas in the simulation. We stress that $M_{\rm cl}$ is in the high-end of a typical GMC mass spectrum \citep{blitz07}. As the clouds in our simulations are submerged in massive HI envelopes, the largest clouds are closer in mass to that of GMC complexes, GMAs (Giant Molecular Associations) or super-clouds \citep{rosolowsky07}. A more realistic analysis should include the full spectrum of cloud masses and their radial distribution but even this simple analysis renders a good agreement with the measured dispersions. The weak dependence on $\kappa$ can explain why most non star-bursting galaxies seem to plateau at a velocity dispersion between 7 and $11\kms$ (see Fig.\,\ref{fig:dib2006} and discussion in Sect.\,\ref{sect:intro}). The rotational velocity varies from $\sim 100\kms$ to $\sim 300\kms$ for most spirals. By assuming a flat rotation curve, $\kappa\propto v_{\rm c}$, the actual change in cloud velocity dispersion between the two limits is, assuming an invariant cloud spectrum, only by a factor of $3^{1/3}$ ($\approx 44\,\%$). 

To conclude, the full picture of gravity driven turbulence in the ISM is based on the existence of dense clouds and filamentary structures all adding to the turbulence budget. Marginally stable gas and cloud-induced filaments generate turbulence through gravitational torques from the swing-amplifier.  Clouds also scatter gravitationally, perturbing the local velocity field even further as well as shock-heating the ISM. The source of turbulence in both cases is self-gravity coupled with shear that in turn converts ordered circular motion of the gas to random velocities, hence tapping rotational energy from the disc. In a sense, self-gravity can here be regarded as a form of viscosity \citep{jogostriker88}. We argue that the mechanisms described in this section serve as a baseline level of turbulence for galaxies where any excess observed velocity dispersion is caused by additional sources such as supernovae activity or magnetic fields coupled with shear. Finally, we note that many of the disc characteristics, e.g. $\sigma$ and $Q_{\rm g}$, show signs of reaching a statistical equilibrium state at late times in the simulations. This is an indication that there is a constant supply rate of energy to the system, here coming from galactic rotation, maintaining a constant level of turbulence. However, on close inspection these quantities are slowly decaying (e.g. the mean velocity dispersion is slowly decreasing, see Fig.\,\ref{fig:veldisp}) due to star formation depleting the disc of gas. Inclusion of realistic gas accretion would affect the temporal evolution of the latter quantities.

\subsubsection{Varying the baryon fraction, shearing motions and cooling floor.}
\begin{figure}
\center
\psfig{file=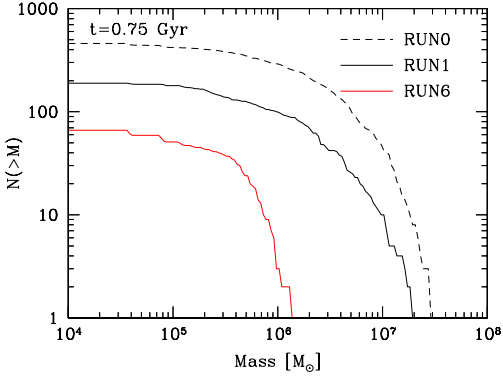,width=200pt}
\caption[]{Cumulative mass spectrum of clouds in RUN0 (dashed line), RUN1 (black) and RUN6 (red) at $t=0.75\Gyr$. The smaller region of instability in the less dense disc in RUN6 brings down the total number of clouds and the mass offset is related to a decrease of the largest wavelength to be stabilized by shear, see text for discussion.}
\label{fig:massspec}
\end{figure}
\begin{figure}
\center
\psfig{file=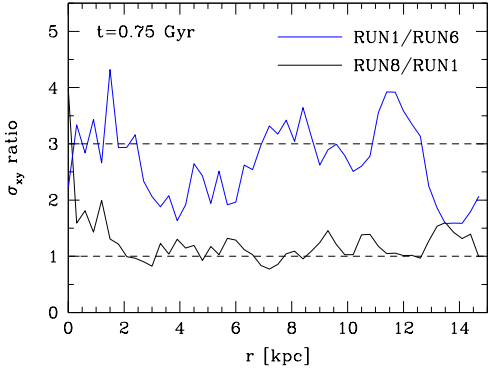,width=200pt}
\caption[]{Ratios of planar velocity dispersions for RUN1, RUN6 and RUN8. The RUN1/RUN6 ratio illustrates the impact of the surface density of the gas, where the more massive disc in RUN1 has $\sim 3$ times larger velocity dispersion. The increased shear in RUN8 increases the velocity dispersion when compared to RUN1, but only significantly in the central parts of the disc.}
\label{fig:dispcomp}
\end{figure}
\begin{figure}
\center
\psfig{file=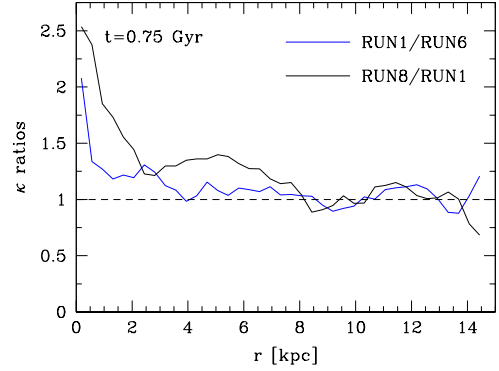,width=200pt}
\caption[]{Ratios of epicyclic frequencies for RUN1, RUN6 and RUN8. The RUN1/RUN6 ratio is above unity in the central parts owing to excess mass transport to the center in RUN1 and hence an increase in central shear. The overall increase in RUN8, where the central parts of the disc are more affected, is expected from the change in dark matter halo concentration.}
\label{fig:kappacomp}
\end{figure}
\begin{figure}
\center
\psfig{file=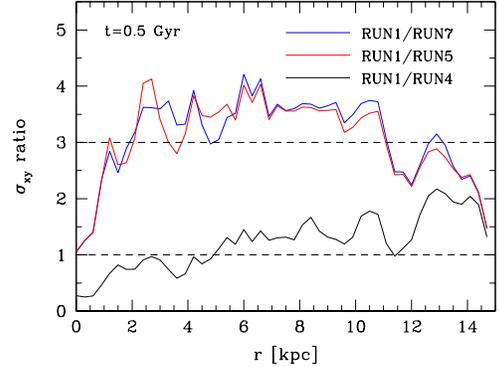,width=200pt}
\caption[]{Ratios of planar velocity dispersions for RUN1, RUN4, RUN5 and RUN7 showing the effect of a temperature floor for the gas cooling. The velocity dispersion in RUN5 and RUN7 are lower by a factor of 3-4 compared to RUN1 due to the inability to undergo gravitational instability. The RUN4 disc is allowed to cool enough to capture the largest clouds forming, hence having similar velocity dispersions as RUN1.}
\label{fig:dispcool}
\end{figure}
In order to explore the effect of the gas density in the initial disc, we keep all parameters fixed apart from the disc mass and carry out an additional simulation (RUN6) in which the gas mass is one third of our fiducial value. This changes the surface density profile and thus the strength of self-gravity. This disc is extremely light and is unstable only within $\sim 8\kpc$, leading to an abundance and mass spectrum of cold gas clouds that is significantly smaller than in RUN0 or RUN1 as seen in Fig.\,\ref{fig:massspec}. Clouds are here defined as isolated clumps of gas satisfying $n>100\,{\rm cm}^{-3}$. The vertical shift, when comparing RUN0 and RUN6, is mostly due to the larger area of instability in the latter simulation and the RUN0 and RUN1 off-set is simply due to star-formation acting to deplete the clouds of high-density gas. We can quantify the observed mass-shift using the arguments presented by e.g. \cite{escala08} where the high-mass end of the spectrum is set by the largest modes not to be stabilized by shear, i.e. the maximum cloud mass will be set by
\begin{equation}
\label{eq:maxcloud}
M_{\rm cl}^{\rm max}=\frac{\pi^4G^2\Sigma_{\rm gas}^3}{4\Omega^4}.
\end{equation}
A factor of 3 decrease in $\Sigma_{\rm gas}$ therefor leads to $M_{\rm cl}^{\rm max}$ being 27 times smaller which is in excellent agreement with the mass spectrum shift in Fig.\,\ref{fig:massspec}. The most massive clouds in RUN6 are $\sim 10^6M_\odot$, leading to weaker cloud-cloud encounters and swing amplified waves, directly lowering the accelerations imparted on nearby gas parcels. This leads to a lower overall velocity dispersions by a factor proportional to $M_{\rm cl}^{1/3}\sim\Sigma$ (from Eq.\, \ref{eq:sigmagammie} and Eq.\,\ref{eq:maxcloud}). An off-set by a factor of $\sim 3$ is confirmed in Fig.\,\ref{fig:dispcomp} which shows the ratios of $\sigma_{xy}(r)$ of the different simulations.

To isolate the effect of the shearing motions we keep all parameters the same as our fiducial RUN1, apart from the rotation curve set by the dark matter halo which is constructed to be much flatter (RUN8). This is achieved by increasing the concentration parameter to $c=40$ whilst maintaining the peak circular velocity to be similar to the rest of our simulations. The swing instability and turbulence induced by shearing motions should be stronger since the epicyclic frequency, defined as
\begin{equation}
\kappa^2=[r\frac{d\Omega^2}{dr}+4\Omega^2],
\label{eq:kappa}
\end{equation}
where $\Omega=v_{\rm c}/r$, increases. For a $v_{\rm c}\sim\sqrt{r}$ this mean that $\kappa\sim1/\sqrt{r}$ and for $v_{\rm c}\sim\,$constant, $\kappa\sim 1/r$. The former $v_{\rm c}$ roughly describes RUN1 and the latter RUN8. Fig.\,\ref{fig:kappacomp} shows $\kappa(r)$ for RUN1, RUN6 and RUN8 at $t=0.75\Gyr$. RUN1 and RUN6 show similar values apart from the inner part of the disc where turbulent viscosity in RUN1 has dragged in more mass compared to RUN6, rendering a more active region. RUN8 shows a steeper behavior, having a $\kappa$ up to 2.5 times larger in the very center of the disc down to a ratio of unity at $r\sim8\kpc$. The larger shear causes a larger $\sigma_{xy}(r)$ compared to RUN1, as seen in Fig.\,\ref{fig:dispcomp}. The effect is strong in the central parts of the disc ($1.5-2$ times larger) and for $r>2\kpc$ the ratio is closer to unity.

It seems evident that gravitational instability is the important driver of velocity dispersions. To explore this further, we artificially truncate the ability to cool gas at different temperatures, hence effectively setting a floor for $Q_{\rm g}$ . In RUN4, RUN5 and RUN7 we introduce a cooling floor for the gas at $10^3$, $10^4$ and $5\,000$ K respectively. In Fig. \ref{fig:dispcool} we compare the ratios of $\sigma_{xy}(r)$ at $t=0.5\Gyr$ and it is evident that RUN5 and RUN7 only show small turbulent motions while RUN4 essentially is as turbulent as the fiducial RUN1. The physical parameters are the same in all simulations except for the temperature of the gas. As $Q_{\rm g}\sim\sigma\sim \sqrt{T}$ it is straightforward to see that RUN5 and RUN7, based on the initial $Q_{\rm g}\sim 2-3$ in Fig.\,\ref{fig:IC}, never acquires $Q_{\rm g}<0.676$. Both simulations only show a weak development of spiral structure. We note that it is plausible that a stellar component would increase the turbulent motions of the discs, as found by \cite{kimostriker07}. Also, the effective observable $\sigma_{z,{\rm eff}}$ in RUN5 and RUN7 is dominated by the thermal component as is at large radii close the values found in RUN1. The cooling in RUN4 can lower $Q_{\rm g}$ by a factor $\sim 3.2$, hence bringing $Q_{\rm g}< 0.676$ while being assisted by non-axissymmetric instabilities. While the smallest scales to be unstable differ in RUN1 and RUN4 due to different pressure support, the larger unstable wavelengths are the same, only limited by the same amount of shear. The presence of the larger, and more dynamically important, clouds and gravitational instabilities makes the velocity dispersion ratio closer to unity.

We conclude that the mass, and hence the surface density, of the disc has a significant impact on the generated turbulent velocity field and we roughly find that, provided the disc is gravitationally unstable, $\sigma\sim\Sigma$. We associate this to the weaker gravitational instabilities present in the disc, which can be seen from the cloud spectrum. The shear of the disc also affects the magnitude of the velocity dispersions but in a weaker fashion. This less strong effect of shear might originate from the $\sim\kappa^{1/3}$-relation (Eq.\,\ref{eq:sigmagammie}) for cloud velocity dispersion. 

\subsubsection{Effect of supernovae feedback}
\label{sect:feedback}
\begin{figure}
\center 
\psfig{file=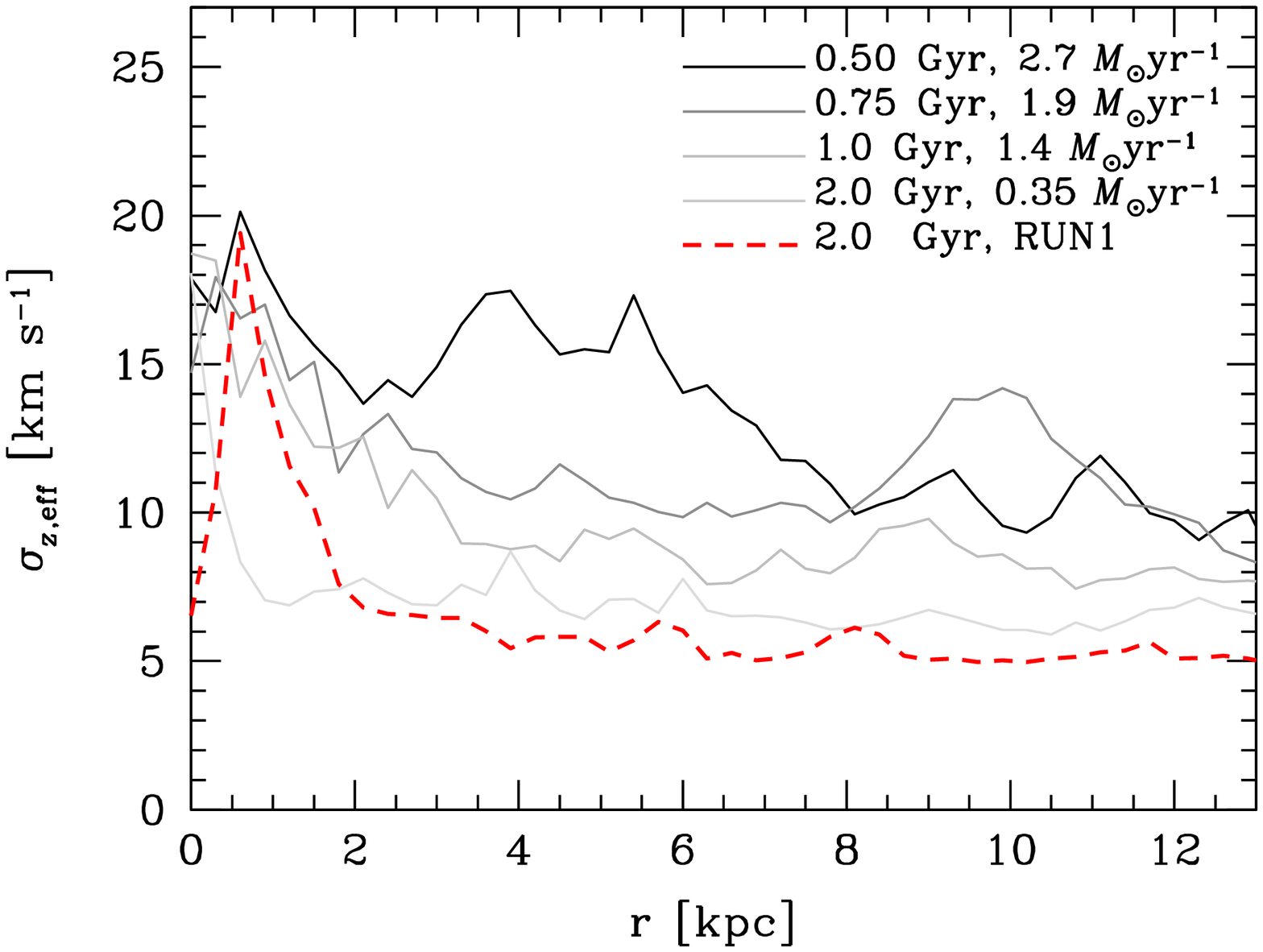,width=200pt}
\caption[]{Effect of star formation rate on the observed vertical velocity
 dispersion in RUN3. The different lines indicate the values at
 different times and therefore also for different SFRs. There is a clear
 trend that a lower stellar activity lowers the measured dispersion,
 approaching the baseline observed dispersion given by RUN1 at
 $t=2.0\Gyr$ (thick red dashed line).}
\label{fig:dispSFR}
\end{figure}
\begin{figure}
\center
\psfig{file=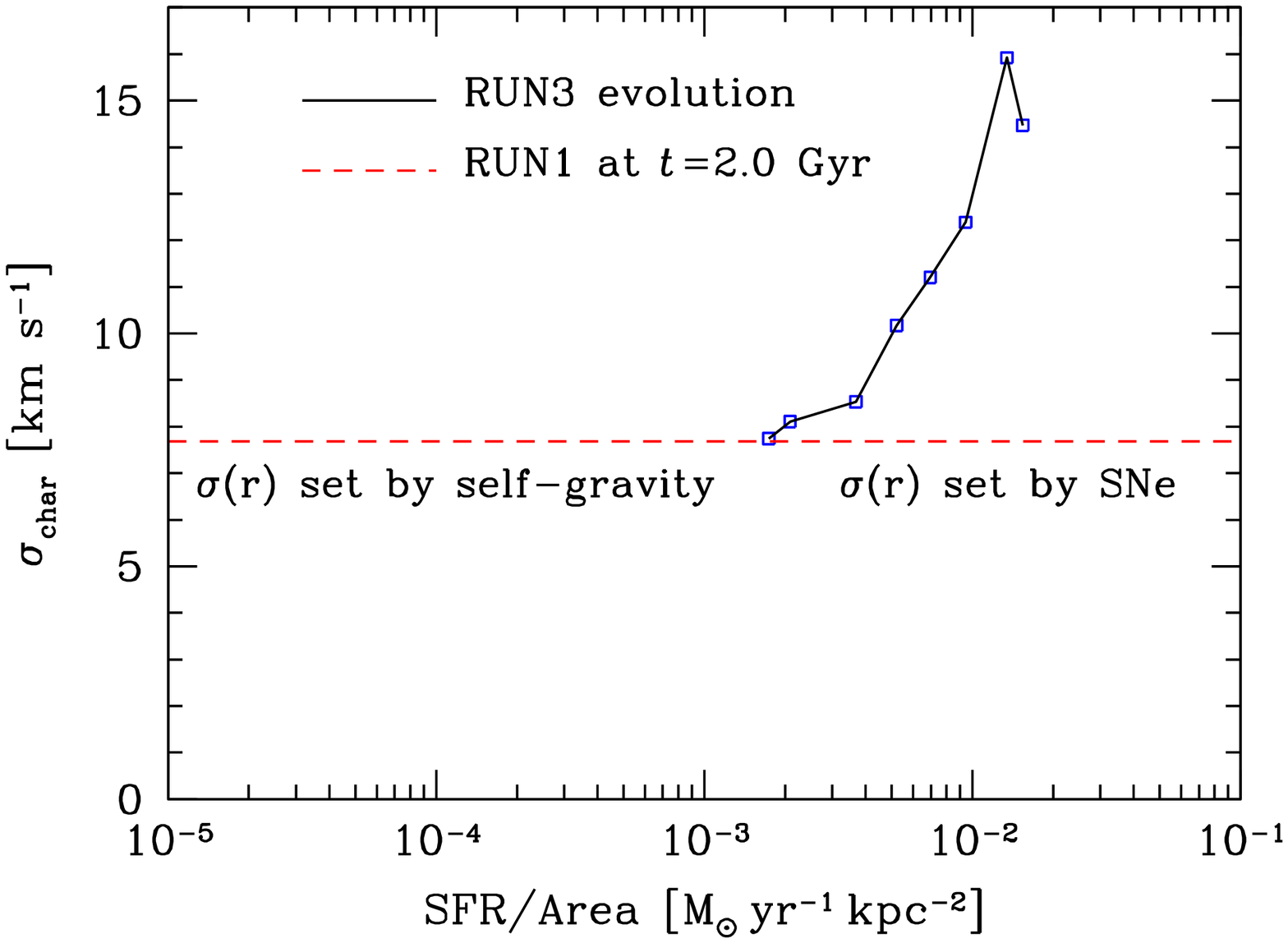,width=200pt}
\caption[]{Effect of star formation rate on the observed vertical velocity
 dispersion in RUN3. The different lines indicate the values at
 different times and therefore also for different SFRs. There is a clear
 trend that a lower stellar activity lowers the measured dispersion,
 approaching the baseline observed dispersion given by RUN1 at
 $t=2.0\Gyr$ (thick red dashed line).}
\label{fig:dispSFR2}
\end{figure}
Self-gravity driven turbulence may be important for galaxies with a low SFR/Area but cannot be the dominant driver behind the large velocity dispersions correlated with high star formation rates. Observations suggests that galaxies with a SFR/Area $\geq {\rm few} \times 10^{-3} M_\odot\,\rm{yr}^{-1}\kpc^{-2}$ show velocity dispersions of several $10 \kms$, see Fig.\,\ref{fig:dib2006}. \cite{dib06} showed that strong SN feedback could explain the transition into this range but were unable to explain the other end of the spectrum. It is plausible that the reason for this stems from their local shearing box approximation that does not take the full disc dynamics into account. Furthermore, \cite{dib06} demonstrated that supernovae driven turbulence is sensitive to poorly known parameters such as efficiency, mass loading, timing etc. Due to computational cost, we can only study one set of parameters. However, as the star formation rate decreases with time (see Fig.\,\ref{fig:SFR}) we are able to study its correlation with velocity dispersion. In Fig.\,\ref{fig:dispSFR} we plot the effective vertical dispersions for RUN3 at different times and hence different SFRs. The general amplitude of the dispersion declines with SFR and after $t=1.5$ Gyr (SFR $\approx 0.74\,M_\odot {\rm yr}^{-1}$) there is little discrepancy between RUN1 and RUN3 suggesting that the effect of SN feedback has saturated. As seen in Fig\,\ref{fig:Tvol}, more warm gas (close to $\sim 10\,000\,{\rm K}$) exists in RUN3 explaining the $\sim 1\kms$ off-set between RUN1 and RUN3 at large radii. This difference is also seen by inspection of the cyan contours in Fig.\,\ref{fig:contoursigma}.

The data shown in Fig.\,\ref{fig:dib2006} can be reproduced by averaging the velocity dispersion and SFR over a suitable area, hence obtaining a characteristic velocity dispersion $\sigma_{\rm{char}}(r)$. \cite{dib06} used the area $A=\pi(3r_0)^2$, where $r_0$ is the scale radius of the stellar disc. Fig.\,\ref{fig:dispSFR2} shows the outcome of this procedure and we clearly detect a supernovae saturation to occur at a SFR/Area of $1-2\times 10^{-3}\,M_\odot\,\rm{yr}^{-1}\kpc^{-2}$, where $\sigma_{\rm char}$ for RUN1 and RUN3 coincides at $t=2.0\Gyr$. This transition from supernovae to self-gravity induced turbulence can explain why the velocity dispersion of NGC 1058 presented in Sect.\,\ref{sect:veldisp} is in good agreement with our simulated disc. NGC 1058 has a derived SFR $\approx 3.5\times 10^{-2}\,M_\odot\rm{yr}^{-1}$ \citep{petric07} which sets the SFR/Area well below $10^{-3} M_\odot {\rm yr}^{-1}\kpc^{-2}$ and hence into the regime where self-gravity induced turbulence can explain the observations. 
\section{Conclusions and discussion}
\label{sect:conclusions}
Three-dimensional, high-resolution hydrodynamical simulations using realistic modelling of star formation and evolution, show that a turbulent ISM naturally develops due to the coupling between gravitational instability and shearing motions. A multiphase medium develops in which cold dense clouds and filaments co-exists with a diffuse warm gas. When supernovae feedback is implemented, a hot phase is present. The marginally stable gas undergoes swing-amplification which both acts to amplify the local density as well as inducing gravitational torques. Cold and dense clouds undergo gravitational scattering, merging and tidal encounters. They also induce waves and filaments in the more diffuse gas which pumps energy into the turbulent process.  The former mechanism stirs the gas even further and we note that the velocity dispersion of the clouds is a fairly good tracer of the HI velocity dispersion.

We summarize our main conclusions here:
\begin{itemize}
\item{Gravitational instabilities in galactic discs leads to a population of massive cold clouds that undergo mutual gravitational interactions and merging. This cloud-cloud harassment process strips material and stirs the ISM. Both cloud interaction and the global non-axisymmetric instability of the disc create non-circular motions from initial ordered rotation. Waves and filaments are generated in the ISM which in turn swing-amplifies to generate further turbulent motions.}

\item{Below a star-formation rate per unit area of $10^{-3}M_\odot {\rm yr}^{-1}$
we find that gravity alone can provide the energy source for maintaining the observed level of turbulence in the ISM of galaxies. The turbulent velocities in our M33 model galaxy have a mean value of $\sim 10\kms$. By calculating an observable HI velocity dispersion, i.e. the contribution from both the turbulent and thermal components, we show that both the magnitude and radial profile is in good agreement of high-resolution HI surveys of e.g. NGC 1058 \citep{dickey90,petric07}. In addition, we reproduce the observed patchy velocity dispersion map.}

\item{Once the star-formation rate exceeds this value, supernovae feedback becomes the dominant driver of turbulence and the velocity dispersion increases with the star-formation rate. This agrees well with the general trend found by \cite{dib06}.}

\item{Lowering the initial gas density weakens the strength of gravitational instability and lowers the resulting cloud mass spectrum, which in turn leads to a lower disc velocity dispersion by a factor $\propto M_{\rm cl}^{1/3}$, as expected from a model in which self-gravity generates significant turbulent motions.}

\item{A direct prediction of this scenario is that galaxies with lower gas fractions at a fixed halo mass should have lower velocity dispersions and different mass fractions in cold, warm and hot phases. Although, detecting the dependence on surface density is complicated by the fact that lower mass galaxies have a higher gas fraction in their discs \citep{mcgaugh05}. In addition, the reaction in low-mass systems to mild stellar activity has not been tested in this work and the outcoming HI velocity dispersion might conspire to render the plateau in Fig.\,\ref{fig:dib2006}.}
\end{itemize}

It is important to note that these results do not rule out the importance of other contributing mechanisms such as supernova feedback or MHD processes, but underscore that self-gravity alone is an important, non-negligible source of turbulence in galactic discs. We believe that this work is complementary to alternative sources of turbulence, see Sect.\,\ref{sect:intro}. For example, \cite{HennebelleAudit07} considered turbulence driven by colliding flows in thermally unstable gas on very small (parsec) scales which are far from resolved in our simulations as we have aimed to resolve the large scale contribution from self-gravity that still would be within the large beam size ($\sim700\pc$). 

Other studies of large scale galactic turbulence includes \cite{wada02} and \cite{wada07} who used an Eulerian code to simulate the dense central part of a galactic disc, where the cold molecular gas phase is dominating. Their results are in agreement with that found here, showing a complicated ISM with a wide range of $Q$-values. Using SPH, \cite{gerritsenicke97} studied star formation and global evolution of the gas in a disc similar to NGC 6503. They demonstrated that a transient flocculant spiral structure with cold cloud complexes is naturally produced in the cold gas, in agreement with our results. The larger amount of warm gas was attributed to heating from stellar photons which is neglected in our work. The subsequent work by \cite{bottema03} extended parameter space to understand the relationship between disc mass and global spiral structure and pointed out the success of swing amplification in predicting this. The measured gas velocity dispersion is similar to that obtained in this paper but was attributed to mechanical forcing from supernovae feedback.

Future work attempting more realistic formation of molecular clouds requires higher resolution and more sophisticated modelling of radiative physics and feedback in order to recover their full range of sizes, masses and life-times, which are affected by internal turbulence and strong feedback disruption.  Even if the actual life and reformation times change, we believe that the global evolution of self-gravity driven turbulence will remain intact as it is not the absolute small scale state of the gas that governs the drag of the diffuse gas but the existence of massive interacting agglomerations. These massive clouds form through gravitational instability that requires seed fluctuations that may be triggered initially by numerical noise, but due to their rapid growth we expect the long term statistical behaviour to be representative. Similarly, our treatment of feedback is quite simplistic and could affect the lifetimes of the smaller mass clouds at the limit of our resolution, cf. end-state of RUN1 and RUN3 in Fig.\,\ref{fig:maps}, although we expect the larger clouds to be stable against these effects. Our initial conditions represent nearby well observed Sc galaxies such as M33 or NGC 1058. As we initialize the baryonic component as gas only we form very massive clouds at early times. However, these structure are significantly reduced in mass at late times due to star formation. 

It is important to point out that the performed simulations do not include an old stellar population in the initial condition. This might change the global evolution to some extent and render a more pronounced spiral structure such as an $m=2$ mode. Such a setup is rather complex and would involve a much larger parameter study. We have postponed this to a further study. In addition, we do not include a background UV field (far and local field). This will change the heating/cooling budget to some extent and can affect the gas density distribution. The global dynamics should however remain the same \citep[see e.g.][]{wada02}.

\section*{Acknowledgments}
O. Agertz would like to thank Andreas Burkert and Woong-Tae Kim for valuable discussion. We thank Mordecai-Mark Mac Low for valuable comments. We thank Doug Potter for making it possible to run the simulations on the zBox2 and zBox3 supercomputers (http://www.zbox2.org) at the University of Z\"urich.

\bibliographystyle{mn2e}
\bibliography{disc.bbl}
\end{document}